# Fast Hamiltonicity checking via bases of perfect matchings

Marek Cygan  
University of Warsaw  
Poland  
cygan@mimuw.edu.pl

Stefan Kratsch[*]  
Max-Planck-Institute  
Saarbrücken, Germany  
skratsch@mpi-inf.mpg.de

Jesper Nederlof[†]  
Utrecht University  
The Netherlands  
J.Nederlof@uu.nl

November 8, 2012

**Abstract**

For an even integer $t \geq 2$, the Matching Connectivity matrix $\mathcal{H}_t$ is a matrix that has rows and columns both labeled by all perfect matchings of the complete graph $K_t$ on $t$ vertices; an entry $\mathcal{H}_t[M_1, M_2]$ is 1 if $M_1 \cup M_2$ is a Hamiltonian cycle and 0 otherwise. Motivated by the computational study of the Hamiltonicity problem, we present three results on the structure of $\mathcal{H}_t$: We first show that $\mathcal{H}_t$ has rank at most $2^{t/2-1}$ over GF(2) via an appropriate factorization that explicitly provides families of matchings $\mathbf{X}_t$ forming bases for $\mathcal{H}_t$. Second, we show how to quickly change representation between such bases. Third, we notice that the sets of matchings $\mathbf{X}_t$ induce permutation matrices within $\mathcal{H}_t$.

Subsequently, we use the factorization to obtain an $1.888^n n^{\mathcal{O}(1)}$ time Monte Carlo algorithm that solves the Hamiltonicity problem in directed bipartite graphs. Our algorithm as well counts the number of Hamiltonian cycles modulo two in directed bipartite or undirected graphs in the same time bound. Moreover, we use the fast basis change algorithm from the second result to present a Monte Carlo algorithm that given an undirected graph on $n$ vertices along with a path decomposition of width at most pw decides Hamiltonicity in $(2+\sqrt{2})^{\text{pw}} n^{\mathcal{O}(1)}$ time. Finally, we use the third result to show that for every $\epsilon > 0$ this cannot be improved to $(2+\sqrt{2}-\epsilon)^{\text{pw}} n^{\mathcal{O}(1)}$ time unless the Strong Exponential Time Hypothesis fails, i.e., a faster algorithm for this problem would imply the breakthrough result of a $(2-\epsilon)^n$ time algorithm for CNF-Sat.

## 1 Introduction

The Hamiltonicity problem and its generalization to the traveling salesman problem are widely acknowledged to be two of the most famous NP-complete problems. Many classical algorithms were invented to tackle these problems (and variants thereof), among them Cristofides' approximation algorithm [12], the Lin-Kernighan heuristic [31], and a polynomial time approximation scheme for Euclidean TSP [1].

A very early and classical result belonging to this list is due to Bellman [4, 5] and, independently, Held and Karp [23]; it demonstrates that the traveling salesman problem can be solved in $\mathcal{O}(n^2 2^n)$ time, where $n$ denotes the number of vertices of the input graph. In order to get this result they introduced dynamic programming over subsets, which became a fundamental design paradigm for obtaining exact exponential time algorithms (see for example [17] and [20, Chapter 3]). Considerable effort has been taken to improve these algorithms: More space efficient algorithms were already given independently by several authors [3, 27, 30].

[*]Part of the work was done at Utrecht University supported by the Nederlandse Organisatie voor Wetenschappelijk Onderzoek (NWO), project: 'KERNELS'.

[†]Supported by the Nederlandse Organisatie voor Wetenschappelijk Onderzoek (NWO), project: 'Space and Time Efficient Structural Improvements of Dynamic Programming Algorithms'.

In his influential survey, Woeginger [38] brought renewed attention to a question suggesting itself already from the 60's: Can either the traveling salesman problem or the Hamiltonicity problem be solved in $(2-\epsilon)^n n^{\mathcal{O}(1)}$ for some $\epsilon > 0$? An affirmative answer was given for the special cases of bounded degree [18, 7, 26, 21] and claw-free graphs [11]. In a breakthrough result, Björklund [6] partially resolved the open question by giving an $1.66^n n^{\mathcal{O}(1)}$ time Monte Carlo algorithm deciding undirected Hamiltonicity, using as a first step a $2^{n/2} n^{\mathcal{O}(1)}$ time algorithm for bipartite graphs. Unfortunately, it seems hard to derandomize this result or to extend it to the traveling salesman problem without incurring a pseudo-polynomial dependence on the input weights. The algorithm of [6] instead of keeping track of all the vertices along the so far constructed path (as Bellman, Held and Karp), labels only some vertices and some edges used on the path. Björklund's algorithm uses counting modulo two as a tool for canceling unwanted self-crossing walks, however it relies on the assumption that the Hamiltonian cycles under consideration satisfy particular counting invariants concerning the number of vertices and edges used. For this reason the algorithm of [6], to the best of our knowledge, can not determine the parity of the number of all Hamiltonian cycles.

Recall that the algorithm of [4, 5, 23] uses dynamic programming based on the observation that if we have constructed a path from $v_a$ to $v_b$, then the only additional information needed is the set of vertices used along the way. Hence the essential information required to store for all subpaths is their endpoints and the set of visited vertices, leading to $\mathcal{O}(n2^n)$ table entries. Recall that for designing a fast dynamic programming algorithm, it is essential that one chooses the right strategy of decomposing a candidate solution. For example, if $G$ has a small balanced separator[1] of size $0.1|V|$ it seems impossible to modify the above algorithm in order to exploit this fact.

In this work we study the following way to decompose a Hamiltonian cycle $H$: Consider an arbitrary ordering $e_1, \ldots, e_m$ of the edges $E$ and decompose $H$ for every $i$ into $H_1 = \{e_1, \ldots, e_i\} \cap H$ and $H_2 = \{e_{i+1}, \ldots, e_n\} \cap H$. This suggests another dynamic programming strategy alternative to [4, 5, 23] that was much less well-studied, mainly because it seemed a priori that the essential information needed to store for a set of partial solutions $H_1$ is $2^{\theta(n \lg n)}$: The degrees (zero, one, or two) of each vertex with respect to $H_1$ and the pairing in which vertices of degree one are connected by $H_1$. The informed reader will notice that this way of decomposing solutions is inherent to dynamic programming algorithms that operate on *tree decompositions* or *path decompositions*.

The notions of *pathwidth/treewidth* proved to be an excellent tool for dealing with many NP-hard problems on graphs. In the 1970s and 1980s, several groups of researchers discovered the concept independently. In their fundamental work on graph minors, Robertson and Seymour [37] introduced the notions *path/treewidth* and *path/tree decomposition*, and these became the dominant terminology. A graph having small path/treewidth means informally that it can be decomposed efficiently in a path/tree-like manner. Many problems can be solved using dynamic programming by decomposing a solution according to the given path/tree decomposition. We refer to [8] for more information on these notions.

In early work, such as the influential result of Courcelle [13], algorithms with running times of the type $f(\mathtt{pw})n$ were given[2]. Later it was noticed that for many such algorithms the function $f(\mathtt{pw})$ can be substantially improved, greatly improving the tractability as well. For example for many problems whose solutions can be verified by separately considering its intersection with all neighborhoods of the vertices of the input graph (see [22]), we can obtain $2^{\mathcal{O}(\mathtt{pw})}n^{\mathcal{O}(1)}$ by employing dynamic programming. For example the Independent Set, or equivalently, Vertex Cover, problem can be solved in $2^{\mathtt{pw}}n$ time [28, 34]. From the work of [25] and standard Karp-reduction it follows that this dependence cannot be improved to sub-exponential algorithms unless the Exponential Time Hypothesis fails, i.e., unless CNF-SAT has a subexponential algorithm. In [32] it was shown that under a stronger assumption (the so-called Strong Exponential Time Hypothesis (SETH)), the current algorithms are optimal even with respect to polynomial jumps, that is, problems with current best running time $f(\mathtt{tw})n^{\mathcal{O}(1)}$ cannot be solved in $f(\mathtt{tw})^{1-\epsilon}n^{\mathcal{O}(1)}$ for positive $\epsilon$ where $f(\mathtt{tw})$ is $2^{\mathtt{tw}}$, $3^{\mathtt{tw}}$ for respectively Independent Set and Dominating Set.

---

[1] A vertex set $X$ such after removing $X$, all connected components have size at most $|V|/2$.
[2] Assuming an input graph $G$ on $n$ vertices along with a path decomposition of $G$ width $\mathtt{pw}$ is given.

For the Traveling Salesman problem, or even the Hamiltonian cycle problem, the situation is different since they are not local problems (see for example [22, Section 5]). In [16], the notion of a Catalan structure was introduced to give an algorithm that solves the $k$-path problems on $H$-minor free graphs in $2^{\mathcal{O}(\mathtt{tw})}$ time. To obtain this, their result works for 'pairing-encodable' problems (i.e. problems where the connectivity properties can be encoded by a matching/pairing) since it bounds the number of ways paths can intersect with a part of the tree decomposition. However, in [15] it was shown that along with many other connectivity problems, Hamiltonian cycle can be solved by a Monte Carlo algorithm in $4^{\mathcal{O}(\mathtt{tw})}n^{\mathcal{O}(1)}$ time. For many of these algorithms with running time $f(\mathtt{tw})n^{\mathcal{O}(1)}$, it was shown that an algorithm running in time $f(\mathtt{tw})^{1-\epsilon}n^{\mathcal{O}(1)}$ would violate SETH. Nevertheless, the exact complexity of the Hamiltonian cycle problem remained elusive.

Recently, a superset of the current authors, found a connection of the optimal substructure of a dynamic programming algorithm with the rank of a certain matrix [9]: Consider a matrix $\mathcal{H}$ with rows and columns indexed by partial solutions, with a 1 if and only if the two partial solutions combined give a valid solution, then if we have more than $\mathrm{rk}(\mathcal{H})$ partial solutions, one will be redundant in the sense that we can safely forget it. This leads to deterministic $2^{\mathcal{O}(\mathtt{tw})}n^{\mathcal{O}(1)}$ time algorithms for many connectivity problems, in a sense derandomizing the work of Cygan et al. [15], but the used techniques do not match their runtimes, e.g., the $4^{\mathtt{tw}}n^{\mathcal{O}(1)}$ time for Hamiltonian cycle.

## Our contribution

Inspired by the mentioned result from [9], we study a matrix that we call the *Matching Connectivity matrix*. We present a family of perfect matchings $\mathbf{X}_t$, which we show to be a basis of the Matching Connectivity matrix, therefore we establish its rank. All further results of the paper use the basis $\mathbf{X}_t$ and its properties as its key tool, which we now elaborate on.

**Determining rank of the Matching Connectivity matrix.** For an even integer $t \geq 2$, the Matching Connectivity matrix $\mathcal{H}_t$ is a matrix that has rows and columns both labeled by all perfect matchings of the complete graph $K_t$ on $t$ vertices; an entry $\mathcal{H}_t[M_1, M_2]$ is 1 if $M_1 \cup M_2$ is a Hamiltonian cycle and 0 otherwise. The centerpiece of our work is that $\mathcal{H}_t$ has rank exactly $2^{t/2-1}$ over GF(2). To establish this result, we define an explicit family $\mathbf{X}_t$ of $2^{t/2-1}$ perfect matchings on $K_t$ and show that its columns (or rows) form a basis of $\mathcal{H}_t$ over GF(2): First, essentially by design, each matching $M \in \mathbf{X}_t$ has a unique partner $M' \in \mathbf{X}_t$ such that their union is a Hamiltonian cycle. Thus, the rows and columns labeled by $\mathbf{X}_t$ induce a permutation matrix, which implies the required rank lower bound. Second, we give an explicit factorization of $\mathcal{H}_t$ into a product of two rectangular matrices with inner dimensions indexed by $\mathbf{X}_t$ in Theorem 3.4. This proves that $\mathbf{X}_t$ is indeed a basis, provides an explicit formula for linear combinations, and completes the claimed rank bound (see Section 3).

In [36] a matrix $\mathcal{H}'_t$ is studied that is obtained by restricting $\mathcal{H}_t$ to all perfect matchings of the complete *bipartite* graph with independent sets of size $t/2$. There it is shown that when taken over the reals, the rank of $\mathcal{H}'_t$ is exactly $\binom{t-1}{t/2-1}$ using group representation theory. Using this the authors disprove the original[3] 'log-rank conjecture' in communication complexity since they also showed that the non-deterministic communication complexity of determining whether two such matchings give a Hamiltonian cycle is $\Omega(n \log \log n)$. In [9, Lemma 3.13] a factorization of a matrix that contains $\mathcal{H}_t$ as submatrix into two matrices with inner dimension $2^{t-1}$ was given.

**Exact algorithms for Hamiltonicity.** By exploiting the peculiar form of the basis $\mathbf{X}_n$ we show that the number of distinct subsets of all the matchings of $\mathbf{X}_n$ is $\mathcal{O}(1.888^n)$. Together with Theorem 3.4 this allows us to show deterministic algorithms computing the parity of the number of Hamiltonian cycles in undirected graphs and directed bipartite graphs (section 3.2.2) in $1.888^n n^{\mathcal{O}(1)}$ time. By combining those results with the Isolation Lemma we obtain Monte Carlo algorithms solving the decision version of Hamiltonicity in both undirected and directed bipartite graphs within the same running time.

Even though our algorithm for undirected graphs is slower than the algorithm of Björklund [6], we believe it is of interest as it uses very much different tools and allows solving the problem also in directed bipartite graphs. We would like to recall that solving the Hamiltonicity problem on undirected bipartite graphs in $\mathcal{O}((2-\epsilon)^n)$ time was the first step of Björklund on the way to the algorithm for general undirected graphs. For this reason we believe that studying directed bipartite graphs is justified.

**Algorithm for bounded pathwidth.** Using the set of perfect matchings $\mathbf{X}_t$ we obtain a faster algorithm for solving the Hamiltonian cycle problem via a non-trivial pathwidth dynamic programming routine. Regarding the algorithm, already from the rank bound it can be argued that only $(2+\sqrt{2})^{\mathtt{pw}}$ space is needed to solve Hamiltonian cycle on a graph with given path decomposition of width $\mathtt{pw}$. The key idea is to replace memoization of all partial solutions by storing only fingerprints of groups of solutions that encode how many partial solutions are consistent with a given basis matching. In fact, these numbers are stored only modulo two which requires only one bit per basis matching. To achieve also *time* $(2+\sqrt{2})^{\mathtt{pw}} n^{\mathcal{O}(1)}$ we show in Lemma 4.1 how to efficiently convert these fingerprints from one basis to another, which permits us to perform the needed dynamic programming table computations (in particular insertion of edges into partial solutions); this crucially depends on the structure of $\mathbf{X}_t$. Notably, Lemma 4.1 gives a second proof of the rank upper bound for $\mathcal{H}_t$, but it does so in a more implicit way; in particular it does not provide an explicit factorization (see Section 4).

Let us point out the main differences to the related algorithmic results obtained by Bodlaender et al. [9]: The present faster algorithm for Hamiltonicity parameterized by pathwidth required a new dynamic programming strategy, unlike the results of [9] that speed up existing formulations. Furthermore, we require randomization (for the Isolation Lemma [33]) to guarantee a unique solution, in order for the fingerprinting approach to work. Finally, achieving time $(2+\sqrt{2})^{\mathtt{pw}} n^{\mathcal{O}(1)}$ depends crucially on the structure of our basis matchings $\mathbf{X}_t$ (to allow for Lemma 4.1) and does not appear to follow directly from the rank. (Similarly, the subsequent lower bound requires the existence of a sufficiently large permutation/identity matrix in $\mathcal{H}_t$ and does not follow directly from the rank lower bound.)

**Matching lower bound assuming SETH.** We show that if the running time of our algorithm can be significantly improved, then satisfiability of CNF-Sat formula's of $m$ clauses and $n$ vertices can be determined in $(2-\epsilon)^n m^{\mathcal{O}(1)}$ time. The latter would contradict the *Strong Exponential Time Hypothesis (SETH)* introduced in [25]. Although there is not consensus about it's truth, a number of results have been given assuming SETH [35, 14, 32, 15]. As with previous results, this result should be interpreted that there is a boundary to significantly improving our algorithm, namely finding the so far elusive $(2-\epsilon)^n m^{\mathcal{O}(1)}$ algorithm for CNF-Sat.

---

[3]A weakened version is still a standing open problem in communication complexity [2, Section 13.2].

For the tight runtime lower bound we use our basis as a part of a gadget in a reduction from CNF-Sat. Although the basic setup is similar to previous lower bounds [32, 15] our reduction is different in the sense that we require a very generic gadget (the induced subgraph gadget, discussed in Subsection 5.1). Using this, we can exploit the crucial property that the submatrix of the Matching Connectivity matrix that is induced by the columns and rows of $\mathbf{X}_t$ is a permutation matrix, i.e., each basis matching has a unique partner in $\mathbf{X}_t$ such that their union is a Hamiltonian cycle. Then, in a similar but technically challenging vein to the reductions of [32, 15], choices can be transferred through a series of gadgets (see Section 5).

**Further algorithmic conclusions.** As a corollary of our bounded pathwidth algorithm in Section 4 (Corollary 4.5) we also obtain a $(2+\sqrt{2})^{n/6} n^{\mathcal{O}(1)} = 1.1583^n n^{\mathcal{O}(1)}$ time Monte Carlo algorithm for Hamiltonicity in cubic graphs, which to the best of our knowledge is the fastest known algorithm in this class of graphs.

## 2 Preliminaries

**Graphs.** We use standard graph notation. For a graph $G = (V, E)$ we denote $V(G)$ and $E(G)$ for its vertex and edge set respectively. For $X, Y \subseteq V$ we let $E(X, Y)$ be the set of all edges with one endpoint in $X$ and one in $Y$. A *Hamiltonian cycle* is the edge set of a simple cycle that visits each vertex exactly once. A *cycle cover* is a set of edges $F \subseteq E$ such that each vertex of $G$ is incident with *exactly two* of these edges (i.e., the edges form cycles). In a *partial* cycle cover each vertex is incident with *at most two* edges (i.e., the edges form paths and cycles).

**Perfect matchings.** A perfect matching of a graph is a set of edges such that each vertex is incident with exactly one of them. It is well known that the union of any two perfect matchings in a graph forms a cycle cover of the graph (where some cycles are potentially of length 2). Given some base set $U$, we use $\Pi_2(U)$ for the set of all perfect matchings of $U$ (i.e., if $U$ has no graph structure then all partitions into sets of size two each is included). Borrowing from the partition lattice partially ordered by refinement, we use $M_1 \sqcap M_2 = \{U\}$, for $M_1, M_2 \in \Pi_2(U)$, to express the fact that the union of the two perfect matchings $M_1$ and $M_2$ is a Hamiltonian cycle; for two perfect matchings this is equivalent to getting the trivial partition $\{U\}$ into a single set as the outcome of the meet-operation $\sqcap$. We do not require any further tools or notation from the partition lattice.

**Pathwidth and path decompositions.** A *path decomposition* of a graph $G = (V, E)$ is a path $\mathbb{P}$ in which each node $x$ has an associated set of vertices $B_x \subseteq V$ (called a *bag*) such that $\bigcup B_x = V$ and the following properties hold:

1. For each edge $\{u, v\} \in E(G)$ there is a node $x$ in $\mathbb{P}$ such that $u, v \in B_x$.

2. If $v \in B_x \cap B_y$ then $v \in B_z$ for all nodes $z$ on the (unique) path from $x$ to $y$ in $\mathbb{P}$.

The *pathwidth* of $\mathbb{P}$ is the size of the largest bag minus one, and the pathwidth of a graph $G$ is the minimum pathwidth over all possible path decompositions of $G$. Since our focus here is on dynamic programming over a path decomposition we only mention in passing that the related notion of treewidth can be defined in the same way, except for letting the nodes of the decomposition form a tree instead of a path.

It is common for the presentation of dynamic programming to use path and tree decompositions that adhere to some simplifying properties, in order to make the description easier to follow. The most commonly used notion is that of a nice tree decomposition, introduced by Kloks [29]; the main idea is that adjacent nodes can be assumed to have bags differing by at most one vertex (this can be achieved without increasing the treewidth). For an overview of tree decompositions and dynamic programming on tree decompositions see [10, 24]. In a similar way, but using also the extension of *introduce edge bags* from [15] we define nice path decompositions as follows.

**Definition 2.1** (Nice Path Decomposition). A *nice path decomposition* is a path decomposition where the underlying path of nodes is ordered from left to right (the predecessor of any node is its left neighbor) and in which each bag is of one of the following types:

- **First (leftmost) bag**: the bag associated with the leftmost node $x$ is empty, $B_x = \emptyset$.

- **Introduce vertex bag**: an internal node $x$ of $\mathbb{P}$ with predecessor $y$ such that $B_x = B_y \cup \{v\}$ for some $v \notin B_y$. This bag is said to *introduce $v$*.

- **Introduce edge bag**: an internal node $x$ of $\mathbb{P}$ labeled with an edge $\{u,v\} \in E(G)$ with one predecessor $y$ for which $u, v \in B_x = B_y$. This bag is said to *introduce $uv$*.

- **Forget bag**: an internal node $x$ of $\mathbb{P}$ with one predecessor $y$ for which $B_x = B_y \setminus \{v\}$ for some $v \in B_y$. This bag is said to *forget $v$*.

- **Last (rightmost) bag**: the bag associated with the rightmost node $x$ is empty, $B_x = \emptyset$.

It is easy to verify that any given path decomposition of pathwidth `pw` can be transformed in time $|V(G)|\mathtt{pw}^{\mathcal{O}(1)}$ into a nice path decomposition without increasing the width.

**Further notation.** For two integers $a, b$ we use $a \equiv b$ to indicate that $a$ is even if and only if $b$ is even. We use Iverson's bracket notation: if $p$ is a predicate we let $[p]$ be 1 if $p$ if true and 0 otherwise. If $\omega : U \to \{1, \ldots, N\}$, we shorthand $\omega(S) = \sum_{e \in S} \omega(e)$ for $S \subseteq U$.

## 3 Structure of the Matching Connectivity matrix and applications

The section is outlined as follows: We will first determine the structure of the Matching Connectivity matrix $\mathcal{H}_t$ in Subsection 3.1. More specifically, we give a basis and determine the rank exactly through an explicit matrix factorization. In Subsection 3.2 we will give an application of the matrix factorization to exact exponential algorithms for Hamiltonicity.

### 3.1 Bases and factorization of Matching Connectivity matrix.

Recall that the matrix $\mathcal{H}_t$ has rows and columns both labeled by all perfect matchings of the complete graph $K_t$ on $t$ vertices; an entry $\mathcal{H}_t[M_1, M_2]$ is 1 if $M_1 \cup M_2$ is a Hamiltonian cycle and 0 otherwise. The *dimension* of the matrix is

$$\frac{t!}{(\frac{t}{2})! \cdot 2^{\frac{t}{2}}} \times \frac{t!}{(\frac{t}{2})! \cdot 2^{\frac{t}{2}}}.$$

| Nr. | | 1 | 2 | 3 | 4 | 5 | 6 | 7 | 8 | 9 | 10 | 11 | 12 | 13 | 14 | 15 | LC |
|---|---|---|---|---|---|---|---|---|---|---|---|---|---|---|---|---|---|
| 1 | ⌒⌒⌒ | 0 | 0 | 0 | 0 | 1 | 1 | 0 | 1 | 1 | 1 | 1 | 0 | 1 | 1 | 0 | 1 |
| 2 | | 0 | 0 | 0 | 1 | 0 | 1 | 1 | 1 | 0 | 0 | 1 | 1 | 1 | 0 | 1 | 2 |
| 3 | | 0 | 0 | 0 | 1 | 1 | 0 | 1 | 0 | 1 | 1 | 0 | 1 | 0 | 1 | 1 | 1+2 |
| 4 | | 0 | 1 | 1 | 0 | 0 | 0 | 0 | 1 | 1 | 1 | 0 | 1 | 1 | 0 | 1 | 4 |
| 5 | | 1 | 0 | 1 | 0 | 0 | 0 | 1 | 0 | 1 | 0 | 1 | 1 | 1 | 1 | 0 | 5 |
| 6 | | 1 | 1 | 0 | 0 | 0 | 0 | 1 | 1 | 0 | 1 | 1 | 0 | 0 | 1 | 1 | 4+5 |
| 7 | | 0 | 1 | 1 | 0 | 1 | 1 | 0 | 0 | 0 | 0 | 1 | 1 | 0 | 1 | 1 | 1+4 |
| 8 | | 1 | 1 | 0 | 1 | 0 | 1 | 0 | 0 | 0 | 1 | 0 | 1 | 1 | 1 | 0 | 2+4+5 |
| 9 | | 1 | 0 | 1 | 1 | 1 | 0 | 0 | 0 | 0 | 1 | 1 | 0 | 1 | 0 | 1 | 1+2+5 |
| 10 | | 1 | 0 | 1 | 1 | 0 | 1 | 0 | 1 | 1 | 0 | 0 | 0 | 0 | 1 | 1 | 2+5 |
| 11 | | 1 | 1 | 0 | 0 | 1 | 1 | 1 | 0 | 1 | 0 | 0 | 0 | 1 | 0 | 1 | 1+4+5 |
| 12 | | 0 | 1 | 1 | 1 | 1 | 0 | 1 | 1 | 0 | 0 | 0 | 0 | 1 | 1 | 0 | 1+2+4 |
| 13 | | 1 | 1 | 0 | 1 | 1 | 0 | 0 | 1 | 1 | 0 | 1 | 1 | 0 | 0 | 0 | 1+2+4+5 |
| 14 | | 1 | 0 | 1 | 0 | 1 | 1 | 1 | 1 | 0 | 1 | 0 | 1 | 0 | 0 | 0 | 1+5 |
| 15 | | 0 | 1 | 1 | 1 | 0 | 1 | 1 | 0 | 1 | 1 | 1 | 0 | 0 | 0 | 0 | 2+4 |

Figure 1: The matrix $\mathcal{H}_6$. Letting the baseset be $\{0, \ldots, 5\}$ matching 1 indexing row and column 1 equals $\{\{0,1\}, \{2,3\}, \{4,5\}\}$. The set $\mathbf{X}_t = \{1, 2, 4, 5\}$ from Definition 3.1 is easily seen to be a row basis; the linear combinations are depicted in the last column.

Let us point our some small cases: For $t = 2$ we have only one perfect matching on the two vertices, and with the union of two such matchings being considered as a Hamiltonian cycle (in all other cases, where $t \geq 4$, there cannot be a Hamiltonian cycle if the two perfect matchings have at least one edge in common). For $t = 4$, there are 3 perfect matchings and $\mathcal{H}_3$ is easily seen to be the complement of the $3 \times 3$ identity matrix. The matrix $\mathcal{H}_6$ is a $15 \times 15$ matrix shown in Figure 1.

To prove the exact value of the rank we introduce for each even $t \geq 2$ a family $\mathbf{X}_t$ of $2^{t/2-1}$ perfect matchings with the goal of proving that the corresponding columns (or rows) of $\mathcal{H}_t$ form a basis over GF(2). The definition of $\mathbf{X}_t$ requires the vertices to be ordered, say $0, 1, \ldots, t-1$, and edges in the matchings are very local, i.e., their endpoints are at most at distance three with respect to the ordering. From the structure of the matchings it will be easy to see that they give a lower bound for the rank of $\mathcal{H}_t$: The submatrix of $\mathcal{H}_t$ induced by rows and columns from $\mathbf{X}_t$ is a permutation matrix, which already has rank $2^{t/2-1}$ itself. This property will be of the essence for our lower bound on the runtime of pathwidth-based dynamic programming for the Hamiltonian cycle problem in Section 5.

Getting the matching upper bound is more involved. We obtain this result by giving a concrete factorization of $\mathcal{H}_t$ in terms of two rectangular submatrices of $\mathcal{H}_t$ induced by the rows respectively columns $\mathbf{X}_t$ in $\mathcal{H}_t$; this is done by a rather technical inductive argumentation. This will form the basis of our algorithm from Subsection 3.2.

Let us begin by introducing the families $\mathbf{X}_t$. For a perfect matching $M \in \Pi_2(U)$ we define a function $\alpha_M \colon U \to U$ with $\alpha_M(i) = j$ if and only if $\{i, j\} \in M$, i.e., $\alpha_M$ maps each element of $U$ to its partner in the perfect matching $M$. We shorthand $U_t := \{0, 1, \ldots, t-1\}$.

**Definition 3.1.** Let $\varepsilon$ denote the empty string. We let $X(2,\varepsilon) := \{\{0,1\}\}$ and $\mathbf{X}_2 := \{X(2,\varepsilon)\}$. Let $t \geq 4$ be an even integer and let $a$ be a bit-string of length $\frac{t}{2} - 2$. We define perfect matchings $X(t, a0)$ and $X(t, a1)$ of $U_t = \{0, \ldots, t-1\}$ as follows:

$$X(t, a1) := X(t-2, a) \cup \{\{t-2, t-1\}\},$$
$$X(t, a0) := (X(t-2, a) \setminus \{\{t-3, \alpha(t-3)\}\}) \cup \{\{t-2, \alpha(t-3)\}, \{t-3, t-1\}\}.$$

We shorthand $X(a)$ for $X(2|a| + 2, a)$ since the bitstring $a$ determines the size $t$ of the base set (namely $U_t$). We use $\bar{a}$ to denote the binary complement of a bit-string $a$. Finally, we let $\mathbf{X}_t$ be the set of all perfect matchings $X := X(t, a)$ for any bitstring $a$ of length $\frac{t}{2} - 1$.

Unfortunately, the formal definition is not very enlightening regarding the actual structure of the perfect matchings $X(t, a)$ in $\mathbf{X}_t$. Let us clarify their structure by first showing the matchings for $t = 4$ and $t = 6$; recall that $X(2, \varepsilon) = \{\{0, 1\}\}$.

$$X(4, 1) = \{\{0,1\}, \{2,3\}\} \qquad X(4, 0) = \{\{0,2\}, \{1,3\}\}$$
$$X(6, 11) = \{\{0,1\}, \{2,3\}, \{4,5\}\} \qquad X(6, 10) = \{\{0,1\}, \{2,4\}, \{3,5\}\}$$
$$X(6, 01) = \{\{0,2\}, \{1,3\}, \{4,5\}\} \qquad X(6, 00) = \{\{0,2\}, \{1,4\}, \{3,5\}\}$$

When recursively constructing further perfect matchings for some even integer $t \geq 4$, we always either add another edge on the two new elements (when the new bit is 1) or we replace the last edge (i.e., the one matching $t-3$ to $t-4$ or $t-5$; note that $t-3$ is the last element for $t' = t-2$) by matching its vertices to the two new elements (when the new bit is 0).

Let us explain the intuition behind the bitstrings: Let $t = 6$ and group the elements as $0 \mid 1, 2 \mid 3, 4 \mid 5$. Observe that the matchings $X(t, \cdot)$ are exactly all choices of matching the elements such that each edge connects two elements that have exactly one dividing vertical line between them, i.e., all choices of perfect matchings that match only elements from adjacent groups. Now the first bit in the bitstring determines whether the first edge is $\{0, 1\}$ or $\{0, 2\}$ (these are all possible options for matching 0 under the group restriction). Depending on this either 2 or 1 still needs to be matched to 3 or 4; the latter choice is determined by the second bit. The last edge must always go to element $t - 1$ (i.e., 5 in this example), so there are only $\frac{t}{2} - 1$ bits.

**Proposition 3.2.** *Let $t \geq 2$ be an even integer, and group the elements $\{0, \ldots, t-1\}$ into $0 \mid 1, 2 \mid \ldots \mid t-3, t-2 \mid t-1$. The family $\mathbf{X}_t$ consists of all perfect matchings that match only elements from different, but adjacent groups. There are $2^{t/2-1}$ such perfect matchings.*

Clearly, the presented families of perfect matchings, one for each even integer $t$, have a very particular and symmetric structure. Our aim is to show that the $2^{t/2-1}$ perfect matchings form a basis for the Matching Connectivity matrix $\mathcal{H}_t$. Regarding any two such matchings $X(t, a)$ and $X(t, b)$ it is not hard to show that their union is a Hamiltonian cycle if and only if $a = \bar{b}$.

**Proposition 3.3.** *Let $t \geq 4$ be an even integer and let $a$ and $b$ be bit-strings of length $\frac{t}{2} - 1$. Then $X(t, a) \cup X(t, b)$ is a Hamiltonian cycle of $K_t$, or equivalently, $X(t, a) \sqcap X(t, b) = \{U_t\}$, if and only if $b = \bar{a}$.*

*Proof.* Consider the first position, say $i$, such that $a[i] = b[i]$ (we consider $a$ and $b$ to be indexed from left to right, starting with 0). All earlier positions $j < i$ are hence different, and following the definition of $X(t, \cdot)$ this means that they prescribe exactly opposite choices. E.g., one matching will match 0 to 1 and the other matches it to 2, without loss of generality $\{0, 1\} \in X(a)$ and $\{0, 2\} \in X(b)$. Consequently, the next bit specifies the matching for 2 in $X(a)$ and the matching for 1 in $X(b)$. This pattern continues and effectively we obtain two alternating paths that start from 0 and follow edges from $X(a)$ and $X(b)$ alternatingly. If the paths meet only at $t-1$ (for which there is no bit that allows an alternate choice of matching) then together they give a Hamiltonian cycle. Since we assumed that $a[i] = b[i]$, the paths meet when the bits prescribe that both matchings match to the same element (it can be verified that bit $i$ decides whether the so far unmatched element of $2i-1$ and $2i$ is matched to $2i+1$ or $2i+2$). Thus, we have found a closed cycle in the union of $X(a)$ and $X(b)$ which does not contain all vertices, hence they do not form a Hamiltonian cycle. (Note the special case of $a[0] = b[0]$ which indicates that both matchings contain $\{0, p\}$ for $p \in \{1, 2\}$.) □

From Proposition 3.3 we directly get a rank lower bound of $2^{t/2-1}$ since the submatrix given by all rows and columns of matchings $X(t, \cdot)$ is a permutation matrix of size $2^{t/2-1} \times 2^{t/2-1}$.

Now we state the main theorem of this section. Due to its technicality, the proof is deferred to Section A. The statement of the theorem is equivalent to saying that the Matching Connectivity matrix $\mathcal{H}_t$ can be written as the product of two rectangular submatrices of $\mathcal{H}_t$ whose rows respectively columns are labeled by matchings from $\mathbf{X}_t$. This implies that the set of those rows/columns forms a basis for $\mathcal{H}_t$ and that its rank is $2^{t/2-1}$.

**Theorem 3.4.** *Let $t \geq 2$ be an even integer and let $M_1, M_2 \in \Pi_2(U_t)$. It holds that*

$$[M_1 \sqcap M_2 = \{U_t\}] \equiv \sum_{a \in \{0,1\}^{t/2-1}} [M_1 \sqcap X(t,a) = \{U_t\}] \cdot [M_2 \sqcap X(t,\bar{a}) = \{U_t\}],$$

*where $X(t,a), X(t,\bar{a}) \in \mathbf{X}_t$ according to Definition 3.1. (Each matching in $\mathbf{X}_t$ occurs exactly twice, once as $X(t,a)$ and once as $X(t,\bar{a})$.)*

**Corollary 3.5.** *The rank of the Matching Connectivity matrix $\mathcal{H}_t$ over GF(2) is $2^{t/2-1}$ for all even integer $t \geq 2$.*

*Proof.* It follows from Proposition 3.3 that the rank is at least $2^{t/2-1}$: The Matching Connectivity matrix $\mathcal{H}_t$ contains a $2^{t/2-1} \times 2^{t/2-1}$ submatrix induced by the columns and rows of all perfect matchings in $\mathbf{X}_t$ which is a permutation matrix.

From Theorem 3.4 we immediately get that the rank is at most $2^{t/2-1}$: We can read the theorem statement as a factorization of $\mathcal{H}_t$ as the product of two submatrices of $\mathcal{H}_t$. The first submatrix has rows labeled by all perfect matchings on $U_t$ and columns labeled by basis matchings $X(t,a)$ for lexicographically ordered bitstrings $a$. The second matrix has columns labeled by all perfect matchings of $U_t$ and rows labeled by basis matchings $X(t,\bar{a})$ for lexicographically ordered bitstrings $a$. The rank upper bound follows immediately from the fact that both matchings have rank at most $2^{t/2-1}$ (corresponding to their smaller dimension). □

## 3.2 Exact exponential algorithms for Hamiltonicity

In this section we present Monte Carlo algorithms for solving the Hamiltonian cycle problem in time $\mathcal{O}(1.888^n \text{poly}(n))$ in undirected graphs and directed bipartite graphs. These algorithms are based on further ideas and insights about the families $\mathbf{X}_t$ of perfect matchings, and in particular we greatly rely on Theorem 3.4.

First, we show that to solve the decision version it is enough to solve the problem of computing the parity of the number of Hamiltonian cycles modulo two. The main part of our algorithm lies in the proofs of the following two lemmas (the proofs are provided in Section 3.2.2).

**Lemma 3.6.** *There is an algorithm, which given an undirected graph $G = (V, E)$ together with a weight function $\omega : E \to \{1, \ldots, \omega_{\max}\}$ finds the parity of the number of Hamiltonian cycles of weight $w$ for every $w \in [0, \ldots, n\omega_{\max}]$ in $\mathcal{O}(1.888^n \text{poly}(n + \omega_{\max}))$ time.*

**Lemma 3.7.** *There is an algorithm, which given a directed bipartite graph $G = (V, A)$ together with a weight function $\omega : A \to \{1, \ldots, \omega_{\max}\}$ finds the parity of the number of Hamiltonian cycles of weight $w$ for every $w \in [0, \ldots, n\omega_{\max}]$ in $\mathcal{O}(1.888^n \text{poly}(n + \omega_{\max}))$ time.*

Now, by an application of the Isolation Lemma, we can show that our modulo two counting of solutions suffices to determine (with high probability) whether or not $G$ is Hamiltonian.

**Definition 3.8.** A function $\omega : U \to \mathbb{Z}$ *isolates* a set family $\mathcal{F} \subseteq 2^U$ if there is a unique $S' \in \mathcal{F}$ with $\omega(S') = \min_{S \in \mathcal{F}} \omega(S)$.

**Lemma 3.9** (Isolation Lemma, [33])**.** *Let $\mathcal{F} \subseteq 2^U$ be a set family over a universe $U$ with $|\mathcal{F}| > 0$. For each $u \in U$, choose a weight $\omega(u) \in \{1, 2, \ldots, N\}$ uniformly and independently at random. Then $\text{prob}[\omega \text{ isolates } \mathcal{F}] \geq 1 - |U|/N$.*

**Theorem 3.10.** *There exists a Monte Carlo algorithm solving the Hamiltonian cycle problem in $\mathcal{O}(1.888^n \text{poly}(n))$ time in undirected graphs and directed bipartite graphs.*

*Proof.* Given a graph $G$ (either undirected, or directed bipartite) with $m$ edges (arcs), for each edge (arc) assign an integer weight from the interval $[1, \ldots, 2m]$ uniformly and independently at random. Then we use Lemma 3.6 (Lemma 3.7) to calculate the parity of the number of Hamiltonian cycles of each weight $w \in [0, \ldots, n\omega_{\max}]$. If for some $w$ there is an odd number of Hamiltonian cycles, then our algorithm returns YES, otherwise it returns NO.

The running time of the algorithm follows from the running time of the black-box usage of the parity calculating algorithm. If there is no Hamiltonian cycle in our graph, then our algorithm certainly returns NO. However, if the graph contains at least one Hamiltonian cycle, then by Lemma 3.9 with probability at least $1/2$ our weight function isolates the family of all Hamiltonian cycles of $G$ and consequently for some weight there is an odd number of Hamiltonian cycles and our algorithm returns YES. Therefore we have obtained a Monte Carlo algorithm. □

### 3.2.1 Further uses of the basis matchings

In this section we give two technical lemmas that form the core of our two algorithms, which are based on the families $\mathbf{X}_n$ of perfect matchings introduced in Section 3. First, we show that the number of *subsets* of all matchings in $\mathbf{X}_n$ is bounded by $\mathcal{O}(1.888^n)$ (Lemma 3.11). Second, we show how to compute the number of extensions of basis matchings to Hamiltonian cycles (Lemma 3.14); for this we use dynamic programming over the mentioned subsets of basis matchings.

**Lemma 3.11.** *The set of all subsets of all matchings from the basis $\mathbf{X}_n$ contains $\mathcal{O}(1.888^n)$ distinct matchings.*

*Proof.* Recall that $\mathbf{X}_n = \{X(n, a) \mid a \in \{0, 1\}^{n/2-1}\}$. For an even integer $n$ define:

$$t(n) = |\{S \subseteq X(n, a) \mid a \in \{0, 1\}^{n/2-1}\}|$$
$$t_2(n) = |\{S \subseteq X(n+2, a) \mid a \in \{0, 1\}^{n/2} \wedge S \cap \{\{n, n+1\}, \{n-1, n+1\}\} = \emptyset\}|.$$

Less formally, $t(n)$ is the number of distinct matchings being subsets of the basis, whereas for $t_2(n)$ we consider longer bitstrings, and additionally we assume that the subset does not contain an edge incident to the last element of the universe, i.e., to $n + 1$. In what follows we prove the following inequalities:

$$t(n) \leq 2t(n-2) + t_2(n-2),$$
$$t_2(n) \leq 4t(n-2) + t_2(n-2).$$

The first inequality follows from the case analysis of the last bit of the string $a$ and whether $S$ contains the edge incident with $n-1$ or not (which we refer to as the *last* edge). When we analyze $t(n)$ we have 4 cases:

- the last bit of $a$ is 0, the set $S$ contains the last edge,
- the last bit of $a$ is 1, the set $S$ contains the last edge,
- the last bit of $a$ is 0, the set $S$ does not contain the last edge,
- the last bit of $a$ is 1, the set $S$ does not contain the last edge.

Note that all the subsets $S$ from the first two cases can be upper bounded by $2t(n-2)$, and the from the second two cases by $t_2(n-2)$, which follows directly from the definition of $t_2$. Now we analyze $t_2(n)$, here we have 8 cases to consider, depending on the last two bits of $a$ (4 choices), and whether $S$ contains the edge between $\{n-3, n-2\}$ and $\{n-1, n\}$ (2 choices), which we call the *penultimate* edge (note that the edge incident to $n+1$ is definitely not contained in $S$). When $S$ contains the penultimate edge, then we bound each of the 4 cases depending on the last two bits of $a$ independently by $t(n-2)$. Observe that so far we made no savings and no cases we considered identical. However the key case is when $S$ does not contain the penultimate edge, which means that $S$ contains neither the penultimate nor the last edge. All 4 such cases can be upper bounded by $t_2(n-2)$, as the last bit of $a$ does not matter anymore.

Observe that $t(2) = 2$ and $t_2(2) = 3$. Moreover if we multiply the horizontal vector $(t(n), t_2(n))$ by the matrix $A = \begin{pmatrix} 2 & 4 \\ 1 & 1 \end{pmatrix}$ we obtain $(a,b)$ where $a \geq t(n+2)$ and $b \geq t_2(n+2)$. At the same time $A = BDB^{-1}$, where

$$B = \begin{pmatrix} \frac{1-\sqrt{17}}{2} & \frac{1+\sqrt{17}}{2} \\ 1 & 1 \end{pmatrix}, D = \begin{pmatrix} \frac{3-\sqrt{17}}{2} & 0 \\ 0 & \frac{3+\sqrt{17}}{2} \end{pmatrix}, B^{-1} = \begin{pmatrix} -\frac{1}{\sqrt{17}} & \frac{1+\sqrt{17}}{2\sqrt{17}} \\ \frac{1}{\sqrt{17}} & -\frac{1-\sqrt{17}}{2\sqrt{17}} \end{pmatrix}.$$

Consequently $A^n = BD^nB^{-1}$ where

$$D^n = \begin{pmatrix} \left(\frac{3-\sqrt{17}}{2}\right)^n & 0 \\ 0 & \left(\frac{3+\sqrt{17}}{2}\right)^n \end{pmatrix}$$

therefore $t(n) = \mathcal{O}(\left(\frac{3+\sqrt{17}}{2}\right)^{n/2})$ and the claimed upper bound follows, since matrices $B$ and $B^{-1}$ contain only fixed constants which are hidden inside the $\mathcal{O}$-notation. In fact one can show that our analysis is tight, since we do not overcount any subsets $S$ neither in $t(n)$ nor in $t_2(n)$ in our case analysis. □

**Definition 3.12 (perfect matchings).** For an undirected graph $H$ by $\Pi_2(H)$ we detote the set of perfect matchings in $H$.

Recall that by $\Pi_2(V)$ we denote the set of all matchings in the complete graph on $V$.

**Definition 3.13 (extensions to Hamiltonian cycle).** For an undirected graph $H = (V, E)$, a weight function $\omega : E \to \{0, \ldots, \omega_{\max}\}$, a matching $M \in \Pi_2(V)$ and an integer $w$ by $\text{ext}(M, w, H, \omega)$ we denote the number of perfect matchings in $H$ of weight $w$ which together with $M$ form a Hamiltonian cycle, i.e.,

$$\text{ext}(M, w, H, \omega) = \sum_{M' \in \Pi_2(H), \omega(M')=w} [M' \sqcap M = \{V\}].$$

**Lemma 3.14.** *Given an undirected graph $G = (V, E)$ and a weight function $\omega : E \to \{0, \ldots, \omega_{\max}\}$ one can in $\mathcal{O}(1.888^n \mathrm{poly}(n + \omega_{\max}))$ time compute all the $(n\omega_{\max}+1)2^{n/2-1}$ values $\mathrm{ext}(X(n, a), w, G, \omega)$ for $a \in \{0, 1\}^{n/2-1}$ and $0 \le w \le n\omega_{\max}$.*

*Proof.* For a (not necessarily perfect) matching $M$ in $G$, a vertex $v$ and weight $0 \le w \le n\omega_{\max}$ let $t[M][v][w]$ be the number of walks in the complete graph on $V$ from an arbitrary fixed vertex $v_1$ to the vertex $v$ which (i) contain an even number of edges, (ii) start with an edge of $M$, (iii) alternately contain an edge of $M$ and an edge of $E$, (iv) use each edge of $M$ exactly once (v) the total weight of edges from $E$ equals $w$. Observe that $\mathrm{ext}(M, w, G, \omega) = t[M][v_1][w]$, hence it is enough to describe a dynamic programming routine computing all the values $t[X(n, a)][v_1][w]$ for $a \in \{0, 1\}^{n/2-1}$, $0 \le w \le n\omega_{\max}$.

We use the recursive formula

$$t[M][v][w] = \sum_{\substack{uu' \in M \\ u \in N(v)}} t[M \setminus \{uu'\}][u'][w - \omega(u, v)].$$

We also define a corner case $t[\emptyset][v][w] = [v = v_1 \wedge w = 0]$ for $v \in V$. By Lemma 3.11 the above formulas together with memoization proved the claimed algorithm. □

### 3.2.2 Proofs of Lemmas 3.6 and 3.7

In this section we prove Lemmas 3.6 and 3.7, that is we focus on computing the number of weighted Hamiltonian cycles modulo two.

*Proof of Lemma 3.6.* First, we iterate over pairs of distinct edges $e_1, e_2$ incident to some arbitrary fixed vertex $v_1$. We want to find the parity of the number of Hamiltonian cycles containing both $e_1$ and $e_2$. Note that for a different pair of edges incident to $v_1$ we count different Hamiltonian cycles and each Hamiltonian cycle is counted exactly once. Since we need an additional assumption that $n$ is even, if this is not the case we subdivide the edge $e_1$, resulting in two new edges $e_1', e_1''$ and we set $\omega(e_1') = \omega(e_1)$, $\omega(e_1'') = 0$, where $e_1'$ is incident to $v_1$. We somewhat abuse notation and by $e_1$ denote the edge $e_1'$. Observe that there is a bijection between weighted Hamiltonian cycles before the subdivision and after the subdivision.

Note that any Hamiltonian cycle containing $e_1$ and $e_2$ can be uniquely decomposed into two perfect matchings $M_1, M_2$ in $G$ where $e_1 \in M_1$ and $e_2 \in M_2$. Let us fix a weight $0 \le w \le n\omega_{\max}$. Our goal is to compute the number of pairs $(M_1, M_2)$, where $M_1, M_2$ are perfect matchings in $G$, $e_1 \in M_1$, $e_2 \in M_2$, $\omega(M_1) + \omega(M_2) = w$ and $M_1 \cup M_2$ is a Hamiltonian cycle. Formally, we want to calculate the following sum modulo two

$$\sum_{\substack{M_1 \in \Pi_2(G) \\ e_1 \in M_1}} \sum_{\substack{M_2 \in \Pi_2(G) \\ e_2 \in M_2 \\ \omega(M_1) + \omega(M_2) = w}} [M_1 \sqcap M_2 = \{V\}].$$

By Theorem 3.4 the above formula is equivalent (modulo two) to

$$\sum_{\substack{M_1\in\Pi_2(G)\\e_1\in M_1}}\sum_{\substack{M_2\in\Pi_2(G)\\e_2\in M_2\\\omega(M_1)+\omega(M_2)=w}}\sum_{a\in\{0,1\}^{n/2-1}}[M_1\sqcap X(n,a)=\{V\}]\cdot[M_2\sqcap X(n,\overline{a})=\{V\}]$$

$$=\sum_{a\in\{0,1\}^{n/2-1}}\sum_{\substack{M_1\in\Pi_2(G)\\e_1\in M_1}}[M_1\sqcap X(n,a)=\{V\}]\sum_{\substack{M_2\in\Pi_2(G)\\e_2\in M_2\\\omega(M_1)+\omega(M_2)=w}}[M_2\sqcap X(n,\overline{a})=\{V\}]$$

$$=\sum_{a\in\{0,1\}^{n/2-1}}\sum_{w_1+w_2=w}\operatorname{ext}(X(n,a),w_1,(V,E_1),\omega)\cdot\operatorname{ext}(X(n,\overline{a}),w_2,(V,E_2),\omega).$$

In the above, we first change the summation order, and then for $i = 1, 2$ denote $E_i = (E \setminus E(v_1, V)) \cup \{e_i\}$. Note that all perfect matchings contributing to $\operatorname{ext}(X(n,a), w_1(V, E_1), \omega)$ (and analogously for $E_2$) contain the edge $e_1$, as this is the only edge in the graph $(V, E_1)$ incident to $v_1$.

By Lemma 3.14 we can find all the needed values in the claimed running time, hence the proof of Lemma 3.6 follows. □

In the proof of Lemma 3.7 we have to deal with directed graphs, while still using Lemma 3.14 which can only handle undirected graphs. However as we will show one can exploit the bipartiteness to harness the directedness of the graph.

*Proof of Lemma 3.7.* Let $G = (V_1 \uplus V_2, A)$ be a directed bipartite graph. Clearly we can assume $|V_1| = |V_2|$, since otherwise there are no Hamiltonian cycles in $G$. We create two auxiliary weighted undirected bipartite graphs $G_\ell = (V_1 \uplus V_2, E_\ell, \omega_\ell)$ and $G_r = (V_1 \uplus V_2, E_r, \omega_r)$ where $E_\ell = \{uv : u \in V_1, v \in V_2, (v,u) \in A\}$, $E_r = \{uv : u \in V_1, v \in V_2, (u,v) \in A\}$ and for $u \in V_1$, $v \in V_2$ we have $\omega_r(uv) = \omega(u,v)$ and $\omega_\ell(uv) = \omega(v,u)$. That is we split the arcs of $A$ depending on whether they have their start-point in $V_1$ or $V_2$ and take the two underlying undirected graphs after the split.

Note that each Hamiltonian cycle in $G$ can be uniquely split into two sets of arcs, one of which corresponds to a perfect matching in $G_\ell$ and the other in $G_r$. Moreover any pair of a perfect matching in $G_\ell$ and a perfect matching in $G_r$ together forms a cycle cover in $G$, which might consists of several cycles. Our goal is to consider all pairs of perfect matchings in $G_\ell$ and $G_r$ which together form a Hamiltonian cycle in the underlying undirected graph of $G$, which guarantees that we count exactly the Hamiltonian cycles in the directed bipartite graph $G$. Let us fix an integer $0 \leq w \leq n\omega_{\max}$ and count the number of Hamiltonian cycles of weight $w$ in the graph $G$.

$$\sum_{\substack{M_1\in\Pi_2(G_\ell)\\}}\sum_{\substack{M_2\in\Pi_2(G_r)\\\omega_\ell(M_1)\cup\omega_r(M_2)=w}}[M_1\sqcap M_2=\{V\}]$$

$$(\text{Thm 3.4})\equiv\sum_{\substack{M_1\in\Pi_2(G_\ell)\\}}\sum_{\substack{M_2\in\Pi_2(G_r)\\\omega_\ell(M_1)\cup\omega_r(M_2)=w}}\sum_{a\in\{0,1\}^{n/2-1}}[M_1\sqcap X(n,a)=\{V\}]\cdot[M_2\sqcap X(n,\overline{a})=\{V\}]$$

$$=\sum_{a\in\{0,1\}^{n/2-1}}\sum_{M_1\in\Pi_2(G_\ell)}[M_1\sqcap X(n,a)=\{V\}]\sum_{\substack{M_2\in\Pi_2(G_r)\\\omega_\ell(M_1)\cup\omega_r(M_2)=w}}[M_2\sqcap X(n,\overline{a})=\{V\}]$$

$$=\sum_{a\in\{0,1\}^{n/2-1}}\sum_{w_1+w_2=w}\operatorname{ext}(X(n,a),w_1,G_\ell,\omega_\ell)\cdot\operatorname{ext}(X(n,\overline{a}),w_2,G_r,\omega_r).$$

By Lemma 3.14 the proof of Lemma 3.7 follows. □

# 4 Solving Hamiltonian cycle fast on pathdecompositions

In this section we present an $n^3(2+\sqrt{2})^{\mathtt{pw}}\mathtt{pw}^{\mathcal{O}(1)}$-time algorithm for solving Hamiltonian cycle on a graph $G$ with a given path decomposition of width $\mathtt{pw}$. Recall that partial solutions for Hamiltonian cycle are sets of paths such that all vertices before the current bag are internal in some path, and vertices in the current bag may be endpoints, internal, or unused. It then suffices to remember for each such partition of the current bag, in what way the endpoints are connected into pairs (these arrangements are perfect matchings on the set of endpoints); it is well-known that any further information about the paths is not needed. The downside is that this involves roughly $\mathtt{pw}^{\mathtt{pw}}$ many partial solutions which dominates the runtime.

The key idea for our much faster algorithm is as follows: Instead of storing for all partitions into endpoints, internal, and unused vertices all the possible perfect matchings of the endpoints, we only store, intuitively, a combined "fingerprint" of all matchings together. Indeed, we fix an ordering of the vertices and store for each matching of the resulting family $\mathbf{X}_t$ the *number* of partial solutions that give a single cycle together with this matching. (These matchings abstract away the need of connecting through all so far unused vertices since this is covered by the partitions.) In fact, since our basis works only over GF(2) we count those solutions modulo two. This however still useful since we can essentially ensure the existence of a unique Hamiltonian cycle of *minimum weight* via the Isolation Lemma (and we need to solve a weighted version of our modulo two counting problem).

Given this setup, let us solve the following problem by dynamic programming on a path decomposition: Given a graph $G = (V, E)$ along with a path decomposition of pathwidth $\mathtt{pw}$, and non-negative edge weights $\omega \colon E \to \{1, \ldots, \omega_{\max}\}$. The task is to compute for each $\omega^* \in \{1, \ldots, n \cdot \omega_{\max}\}$ the parity of the number of Hamiltonian cycles of $G$ with weight exactly $\omega^*$. We assume that we are given a nice path decomposition for $G$ of width at most $\mathtt{pw}$; we treat the decomposition as a sequence of bags that are ordered from left to right. To solve this problem we proceed as outlined above: For each partition into internal, endpoints, and unused vertices we take a basis for the perfect matchings on the endpoints and compute (and store) the number of partial solutions that are consistent with each perfect matching. We maintain and process this information throughout the dynamic programming; the main work is spent (unsurprisingly) on bags that introduce edges since this causes a rather involved recomputation of fingerprints (as we cannot work explicitly on separate partial solutions).

For technical convenience our algorithm "guesses" one edge incident on a vertex of degree at most $\mathtt{pw}$ to be used in the Hamiltonian cycle. Given a nice path decomposition it can be easily seen that the rightmost introduce vertex bag can only introduce a vertex $v$ of degree at most $\mathtt{pw}$: all its neighbors must be in the current bag and no additional possible neighbors can be added on the right. It can be easily verified that all remaining bags, namely introduce edge and forget vertex bags, can be reordered freely under the constraint that no vertex is forgotten before all its edges were introduced. Thus, picking any edge incident on $v$, say $\{u, v\}$, we can reorder such that the last bags are: 1) introduce $\{u, v\}$ (with current vertex set $\{u, v\}$, 2) forget vertex $u$, 3) forget vertex $v$. Our computation (from left to right) may then stop at bag 1) and (for some choice $\omega^*$ of total weight) check the parity of the number of partial cycle covers that would form a Hamiltonian cycle of weight $\omega^*$ when augmented with the edge $\{u, v\}$.

Let the vertices of $G = (V, E)$ be ordered arbitrarily, say $V = \{v_1, \ldots, v_n\}$, and let the weight of any edge $\{v_i, v_j\}$ be given by $\omega(v_i, v_j)$. We perform dynamic programming on the given path decomposition, proceeding from left to right (until we reach the introduce edge bag of the "guessed" edge). At each bag, with some vertex set $B$, we compute table entries $t[B_0, B_1, B_2, w, M]$ for all partitions $B = B_0 \cup B_1 \cup B_2$, all integers $\omega \in \{0, \ldots, n \cdot \omega_{\max}\}$, and all perfect matchings $M$ from a basis for $B_1$ (the latter is according to Definition 3.1 with a standard arbitrarily fixed ordering induced from $V$). Each entry contains the partity of the number of partial cycle covers $C$ of the graph induced by all vertices left of and including the current bag and all edges introduced so far, such that

1. $C \cup M$ is a single cycle,

2. the total weight of the edges in $C$ is equal to $\omega$,

3. the vertices in $B_i$ have degree exactly $i$ in $C$,

4. and all vertices that only occur left of the current bag have degree two; we denote those by $B_\ell$.

We call $C$ a $(B_0, B_1, B_2, B_\ell, \omega)$-cycle cover if it respects 2., 3., and 4. If it respects all four properties then we call it a $(B_0, B_1, B_2, B_\ell, \omega, M)$-cycle cover, i.e., if additionally the union with $M$ is a single cycle.

The main technical difficulty in the dynamic programming lies in handling the information stored with respect to the basis for perfect matchings of $B_1$, in particular when introducing a new edge in the path decomposition. It is crucial that we can efficiently compute a representation of the same information with respect to a different ordering. Intuitively, the following lemma allows us to change the basis of our representation. Since we apply the lemma separately for each partition $B = B_0 \dot\cup B_1 \dot\cup B_2$, set of previous vertices $B_\ell$, and choice of weight $\omega$, we state it in terms of a simplified table with one entry $T[M]$ for each basis matching $M$. The lemma applies to our dynamic programming application by letting $\mathcal{C}$ be the set of all $(B_0, B_1, B_2, B_\ell, \omega)$-partial cycle covers and letting $S = B_1$.

**Lemma 4.1.** *Let $\mathcal{C}$ denote an arbitrary set of partial cycle covers and let $\mathbf{X}$ denote a family of basis matchings on some totally ordered set $S = \{v_0, \ldots, v_{t-1}\}$ of vertices of even size. Furthermore, for each $M \in \mathbf{X}$, let $T[M]$ denote the number of partial cycle covers $C \in \mathcal{C}$ such that $M \cup C$ is a single cycle. For any $\mathbf{X}'$ with respect to any other ordering of $S$, the corresponding values $T'[M']$ for $M' \in \mathbf{X}'$ can be computed in time $2^{t/2-1} t^{O(1)}$.*

*Proof.* Clearly, any permutation of the ordering of $S$ can be achieved by at most $t^2$ swaps of two consecutive elements. Thus, it suffices to show how to move some vertex $v$ one step to the "right" by swapping it with its successor in the order. We are able to show that the computation of any value $T'[M']$ requires only the contents of at most three other entries in $T[\cdot]$.

Recall the grouping of $\{v_0, \ldots, v_{t-1}\}$ into $v_0 \mid v_1, v_2 \mid \ldots \mid v_{t-3}, v_{t-2} \mid v_{t-1}$. We have to distinguish a few cases about the position of $v$ (odd or even) and a few of the edges in $M$ that involve elements close to $v$ in the ordering. For brevity we shorthand a little: we use $a \mid b, c \mid d, e \mid f$ to denote a part of the ordering (and its groups), with $a$ matched to $b$ or $c$ and $f$ matched to $d$ or $e$. This includes that we do not specify the positions of $a$ and $f$ in their groups, since it can be checked that they are immaterial for the discussion below (we know that one of the elements in the preceding and one in the subsequent group are matched like that).

If $v \in \{v_1, v_3, \ldots, v_{t-1}\}$ (i.e., odd number and even position), then we have $a \mid v, c \mid d, e \mid f$ (and $v$ cannot be in the first group since that only contains $v_0$). Consider the same ordering but with $c$ and $v$ flipped, i.e., (locally) we have $a \mid c, v \mid d, e \mid f$, and let $M$ be a basis matching for that ordering. It is easy to see that $M$ is also in the basis for the initial ordering, and the corresponding bitstring is obtained by inverting the bit that corresponds to the choice of matching $a$ to $v$ or $c$. (The bit effectively decides whether $a$ is matched to the first or second element of the group $|c, v|$; flipping the order in the group as well as the bit cancels out.)

The case of an odd position is, unfortunately, more involved since $v$ moves to a different group. Let $v$ be in an odd position, i.e., $v \in \{v_0, v_2, v_4, \ldots, v_{t-2}\}$. Thus we have $a \mid b, v \mid d, e \mid f$ (for now let us assume that $v$ is not the first element in the ordering). Moving $v$ one step to the right results in $a \mid b, d \mid v, e \mid f$; let $M$ be some basis matching for this ordering. There are four ways in which $M$ can match the vertices $a, b, d, v, e, f$ taking into account our choice of $a$ and $f$, and they have to be treated differently (the cases are equivalent to the four matchings in the basis $\mathbf{X}_6$ for six vertices).

**i)** If $\{a, b\}, \{d, v\}, \{e, f\} \in M$, then $M$ is also in the basis for the initial ordering, and with the same bitstring. Hence
$$T'[M] \equiv T[M].$$

**ii)** If $\{a, b\}, \{d, e\}, \{v, f\} \in M$, then consider the following two matchings $M_1$ and $M_2$ which are obtained by replacing the matching on $d, e, v, f$ by either of the other two possible options (there are three different perfect matchings on 4 elements).
$$M_1 := (M \setminus \{\{d, e\}, \{v, f\}\}) \cup \{\{d, v\}, \{e, f\}\},$$
$$M_2 := (M \setminus \{\{d, e\}, \{v, f\}\}) \cup \{\{d, f\}, \{e, v\}\}.$$

It is easy to see that $M_1$ and $M_2$ are basis matchings for the initial ordering $a \mid b, v \mid d, e \mid f$, since all their edges connect adjacent groups. Crucially, any partial cycle cover $C \in \mathcal{C}$ gives a single cycle with $M$ if and only if this is true for exactly one of $M_1$ and $M_2$. To see this, it suffices to consider the ways in which $a, b, v, d, e, f$ are connected (the connection through $C$ can be abstracted to simply a perfect matching on these six vertices).
$$T'[M] \equiv T[M_1] + T[M_2]$$

**iii)** If $\{a, d\}, \{b, v\}, \{e, f\} \in M$, then we use the same edge-flipping argument, but on $\{a, d\}$ and $\{b, v\}$. We obtain
$$M_1 := (M \setminus \{\{a, d\}, \{b, v\}\}) \cup \{\{a, b\}, \{d, v\}\},$$
$$M_2 := (M \setminus \{\{a, d\}, \{b, v\}\}) \cup \{\{a, v\}, \{b, d\}\},$$

which are both in the basis with respect to the initial ordering. Thus, the desired table value can be computed as
$$T'[M] \equiv T[M_1] + T[M_2].$$

**iv)** Finally, we have $\{a, d\}, \{b, e\}, \{v, f\}$. In this case we use three basis matchings from the previous ordering, namely
$$M_1 := (M \setminus \{\{a, d\}, \{b, e\}, \{v, f\}\}) \cup \{\{a, b\}, \{d, f\}, \{e, v\}\},$$
$$M_2 := (M \setminus \{\{a, d\}, \{b, e\}, \{v, f\}\}) \cup \{\{a, v\}, \{b, d\}, \{e, f\}\},$$
$$M_3 := (M \setminus \{\{a, d\}, \{b, e\}, \{v, f\}\}) \cup \{\{a, v\}, \{b, e\}, \{d, f\}\}.$$

Again, these matchings are in the basis for the initial ordering since all edges are between adjacent groups. It can be checked that the desired table entry is exactly the sum modulo two of the table entries corresponding to these three matchings. To see this consider Table 1, and identify $(a, d, b, e, v, f)$ with $(0, 1, 2, 3, 4, 5)$. Then $\{a, d\}, \{b, e\}, \{v, f\}$ corresponds to matching 1, $M_1$ to matching 6, $M_2$ to matchings 10 and $M_3$ to matching 12, and the corresponding rows are easily seen to be linearly dependent.
$$T'[M] \equiv T[M_1] + T[M_2] + T[M_3]$$

Since there are $2^{t/2-1}$ perfect matchings in the basis for any ordering of $S$ and each entry takes only $\mathcal{O}(1)$ operations the claimed total time follows by computing all intermediate tables (only the most recent one needs to be stored). □

Now we can return to the description of our algorithm. The dynamic programming on the path decomposition proceeds from "left" to "right", using the table of the previous bag, denoted $t'[]$, to compute the table of the current bag, denoted $t[]$. We use $t'_{uv}$ and $t_{uv}$ for the corresponding tables that are obtained by changing ordering and basis in such a way that $u$ and $v$ are the last two vertices in the total order. For introduce and forget vertex bags there is not much work required since by themselves vertices do not affect our partial solutions. The main work lies in the computations required for the introduce edge bags: Partial solutions that use the new edge have a different set $B_1$ of degree-1 vertices, which comes with a different basis.

**First bag:** For the first bag the vertex set $B$ is empty and we only get a single trivial table entry

$$t[\emptyset, \emptyset, \emptyset, 0, \emptyset] \equiv 0.$$

**Introduce vertex bag:** We have a current bag with vertex set $B$ that introduces a vertex $v$. Accordingly the previous bag has vertex set $B \setminus \{v\}$. Furthermore, the set of vertices that occur only on the left is the same for both bags. Clearly, in the subgraph given by all vertices of the current and preceding bags plus all edges introduced so far no partial cycle cover can include edges incident on $v$. Thus

$$t[B_0, B_1, B_2, \cdot, \cdot] \equiv 0,$$

for all partitions $B = B_0 \dot\cup B_1 \dot\cup B_2$ with $v \in B_1 \cup B_2$. For partitions with $v \in B_0$ and any choice of weight $\omega$ and $B_1$-basis matching $M$ we already have the correct number in the table for the previous bag, namely

$$t[B_0, B_1, B_2, \omega, M] \equiv t'[B_0 \setminus \{v\}, B_1, B_2, \omega, M].$$

**Forget vertex bag:** We have a current bag with vertex set $B$ that "forgets" vertex $v$; the previous bag has vertex set $B \cup \{v\}$. Let $B_\ell$ be the vertices that occur only left of the current bag and note that $B_\ell \setminus \{v\}$ occur only left of the previous bag. Fix any partition $B = B_0 \dot\cup B_1 \dot\cup B_2$, a matching $M$ from the $B_1$-basis (with standard ordering), and a weight $\omega$. If $C$ is a $(B_0, B_1, B_2, B_\ell, \omega, M)$-cycle cover, then $v$ must have degree two in $C$ by definition. Conversely, any $(B_0, B_1, B_2 \cup \{v\}, B_\ell \setminus \{v\}, \omega, M)$-cycle cover $C'$ (whose number modulo two is stored in the previous bag) is also a $(B_0, B_1, B_2, B_\ell, \omega, M)$-cycle cover. Hence

$$t[B_0, B_1, B_2, \omega, M] \equiv t'[B_0, B_1, B_2 \cup \{v\}, \omega, M].$$

**Introduce edge bag:** We have a current bag with vertex set $B$ that introduces an edge $\{u, v\}$ with $u, v \in B$. The previous bag has the same vertex set $B$ (but its partial solutions cannot make use of $\{u, v\}$), and both bags have the same set $B_\ell$ of vertices that occur only left of them. First, we compute from the table $t'[]$ of the previous bag the table $t'_{uv}[]$; i.e., the information is represented with respect to a basis that has $u$ and $v$ as the last two vertices in the ordering using Lemma 4.1. We will proceed by computing $t_{uv}[]$ which will subsequently be transformed to $t[]$; this completes the procedure for an introduce edge bag.

We explain the computation of $t_{uv}[B_0, B_1, B_2, \omega, M]$ from $t'_{uv}[]$ for an arbitrary partition $B = B_0 \dot\cup B_1 \dot\cup B_2$, integer weight $\omega$, and perfect matching $M$; the latter is from the basis for $B_1$ with modified ordering of $V$ ($u$ and $v$ are last two elements). We need to consider different cases depending on which sets $B_i$ contains $u$ and $v$.

**i) $u \in B_0$ or $v \in B_0$:** In this simple case no $(B_0, B_1, B_2, \omega)$-cycle cover may contain the edge $\{u, v\}$ and the same is true for all perfect matchings $M$ on $B_1$ (in particular for those from a basis). Thus, directly from the definition of our tables we get that

$$t_{uv}[B_0, B_1, B_2, \omega, M] \equiv t'_{uv}[B_0, B_1, B_2, \omega, M],$$

for all basis matchings $M$.

**ii) $u, v \in B_1$:** Since $M$ is a perfect matching from a basis on $B_1$ it may contain the edge $\{u, v\}$. If $\{u, v\} \in M$ then no $(B_0, B_1, B_2, B_\ell, \omega, M)$-cycle cover $C$ can contain $\{u, v\}$ (since the two copies of $\{u, v\}$ would already give a separate cycle). (In the final introduce edge bag we do not need to perform this computation since we only check for a solution.) We already know the parity of the number of $(B_0, B_1, B_2, B_\ell, \omega, M)$-cycle covers avoiding the edge $\{u, v\}$ from $t'_{uv}[]$ and get

$$t_{uv}[B_0, B_1, B_2, \omega, M] \equiv t'_{uv}[B_0, B_1, B_2, \omega, M].$$

Otherwise we have $\{u, v\} \notin M$ which implies that $\{p, u\}, \{q, v\} \in M$ for some $p, q \in B_1 \setminus \{u, v\}$ (with $p \neq q$) since $M$ is a perfect matching of $B_1$. As before we already know the parity of the number of $(B_0, B_1, B_2, B_\ell, \omega, M)$-cycle covers that do not contain $\{u, v\}$ from $t'_{uv}[]$.

Consider a $(B_0, B_1, B_2, B_\ell, \omega)$-cycle cover $C$ that does contain $\{u, v\}$. Thus $C \cup M$ contains a cycle $S = (x_1, \ldots, x_r, p, u, v, q, x_1)$. Intuitively, we want to compare this to partial cycle covers without $\{u, v\}$; we effectively move the edge into the matching: Let $C' := C \setminus \{\{u, v\}\}$ and let $M' := (M \setminus \{\{p, u\}, \{q, v\}\}) \cup \{\{p, q\}\}$ (the additional edge in $M$ is modeled by contracting $p, u, v, q$ into just $p, q$). Observe that $C'$ is a $(B_0 \cup \{u, v\}, B_1 \setminus \{u, v\}, B_2, B_\ell, \omega - \omega(u, v))$-cycle cover. Clearly, $C' \cup M'$ contains a cycle $S' = (x_1, \ldots, x_r, p, q, x_1)$, and all further cycles (if they exist) are the same as in $C \cup M$. Thus $C$ is a $(B_0, B_1, B_2, B_\ell, \omega, M)$-cycle cover if and only if $C'$ is a $(B_0 \cup \{u, v\}, B_1 \setminus \{u, v\}, B_2, B_\ell, \omega - \omega(u, v), M')$-cycle cover. This fact is useful for the computation, provided that $M'$ is in the basis for $B_1$ with modified ordering, which will check next.

Let us see that $M'$ is indeed in the basis for $B_1$: First of all, since $u$ and $v$ are at the end of the ordering of $B_1$, when we remove these two vertices the remaining vertices keep their current ordering. Second, due to the structure of bases for $B_1$, we see that the edges $\{p, u\}, \{q, v\} \in M$ imply that the ordering of $B_1$ ends either with $\ldots, p, q, u, v$ or with $\ldots, p, r, q, u, v$ (to see this, recall the grouping of the elements and the fact that the basis matchings are exactly all ways of pairing up elements from adjacent groups). Accordingly the ordering for $B_1 \setminus \{u, v\}$ ends in $\ldots, p, q$ or $\ldots, p, r, q$. If we have $\ldots, p, q, u, v$ then $M = X(|B_1|, a10)$ and $M' = X(|B_1| - 2, a1)$, for some bitstring $a$. If we have $\ldots, p, r, q, u, v$ then $M = X(|B_1|, a00)$ and $M' = X(|B_1| - 2, a0)$. In both cases, $M'$ is part of the basis for $B_1 \setminus \{u, v\}$ (in fact, its corresponding bitstring is the same as the one for $M$ minus the last position). Thus, taking into account the contribution for cycle covers without $\{u, v\}$, we can compute $t_{uv}[B_0, B_1, B_2, \omega, M]$ as follows

$$\begin{aligned}t_{uv}[B_0, B_1, B_2, \omega, M] &\equiv t'_{uv}[B_0, B_1, B_2, \omega, M] \\ &+ t'_{uv}[B_0 \cup \{u, v\}, B_1 \setminus \{u, v\}, B_2, \omega - \omega(u, v), M'].\end{aligned}$$

**iii) $u \in B_1$ and $v \in B_2$:** We get that $\{u, v\} \notin M$, but there must be some $p \in B_1 \setminus \{u\}$ with $\{p, u\} \in M$ since $M$ is a perfect matching of $B_1$. Again we get the contribution of $t'_{uv}[B_0, B_1, B_2, \omega, M]$ for all $(B_0, B_1, B_2, B_\ell, \omega, M)$-cycle covers that do not contain the edge $\{u, v\}$.

If a $(B_0, B_1, B_2, B_\ell, \omega)$-cycle cover $C$ contains the edge $\{u, v\}$, then $C \cup M$ contains a cycle $(x_1, \ldots, x_r, v, u, p, x_1)$. Again, we effectively move the edge $\{u, v\}$ "into" $M$: Consider $C' :=$ $C \setminus \{\{u,v\}\}$ and $M' := (M \setminus \{\{p,u\}\}) \cup \{\{p,v\}\}$. Observe that $C'$ is a $(B_0 \cup \{u\}, (B_1 \setminus \{u\}) \cup \{v\}, B_2 \setminus \{v\}, B_\ell, \omega - \omega(u,v))$-cycle cover. Furthermore, $C' \cup M'$ contains a cycle $(x_1, \ldots, x_r, v, p, x_1)$, and all further cycles (if there are any) are the same as in $C \cup M$. Thus $M \cup C$ is a single cycle if and only if $M' \cup C'$ is a single cycle. Again we need to check that $M'$ is in the basis for $(B_1 \setminus \{u\}) \cup \{v\}$. In the present case this is straightforward: Since $u$ and $v$ are last in the modified ordering of $V$, they both occupy the last position on $B_1$ and $(B_1 \setminus \{u\}) \cup \{v\}$ respectively. Since $M$ and $M'$ differ exactly by replacing $u$ with $v$, we get that $M'$ is in the basis for $(B_1 \setminus \{u\}) \cup \{v\}$ and, in fact, corresponds to the same bitstring as $M$ for $B_1$. We get

$$t_{uv}[B_0, B_1, B_2, \omega, M] \equiv t'_{uv}[B_0, B_1, B_2, \omega, M]$$
$$+ t'_{uv}[B_0 \cup \{u\}, (B_1 \setminus \{u\}) \cup \{v\}, B_2 \setminus \{v\}, \omega - \omega(u,v), M'].$$

**iv)** $u \in B_2$ **and** $v \in B_1$: This case is symmetric to the previous one. Despite $u$ being second to last and $v$ being last in the modified ordering, this requires no change of argumentation since we always have only one of them in the set of matched vertices.

**v)** $u, v \in B_2$: Clearly no perfect matching $M$ on $B_1$ can contain $\{u,v\}$ or any other edge incident with $u$ or $v$. However, $(B_0, B_1, B_2, B_\ell, \omega)$-cycle covers $C$ may contain the edge $\{u,v\}$. The parity of the number of such partial cycle covers that do not use $\{u,v\}$ and that are consistent with $M$ is already stored in $t'_{uv}[B_0, B_1, B_2, \omega, M]$.

Let us consider a $(B_0, B_1, B_2, B_\ell, \omega)$-cycle cover $C$ that does contain $\{u,v\}$. Let $C' := C \setminus \{\{u,v\}\}$ and $M' = M \cup \{\{u,v\}\}$. Observe that $C'$ is a $(B_0, B_1 \cup \{u,v\}, B_2 \setminus \{u,v\}, B_\ell, \omega - \omega(u,v))$-cycle cover. It is easy to see that $C \cup M$ is a single cycle if and only if $C' \cup M'$ is a single cycle, since $C \cup M = C' \cup M'$.

Let us check that $M'$ is in the basis for $B_1 \cup \{u,v\}$ with modified ordering. Let the end of the ordering of $B_1$ be $\ldots, p, q, r$. If $M = X(|B_1|, a0)$ then $\{p, r\} \in M$, and we get $M' = X(|B_1| + 2, a01)$. Else, if $M = X(|B_1|, a1)$ then $\{q, r\} \in M$ and we get $M' = X(|B_1| + 2, a11)$. (Note that we basically only need to append a 1 to the bitstring for $M$ to enforce that $p$ or $q$ is matched to $r$ and not to $u$; the edge $\{u,v\} \in M'$ is then implied.) Thus $M'$ is indeed in the basis for $B_1 \cup \{u,v\}$ with modified ordering. We get

$$t_{uv}[B_0, B_1, B_2, \omega, M] \equiv t'_{uv}[B_0, B_1, B_2, \omega, M]$$
$$+ t'_{uv}[B_0, B_1 \cup \{u,v\}, B_2 \setminus \{u,v\}, \omega - \omega(u,v), M \cup \{\{u,v\}\}].$$

Now we have handled all cases for computing $t_{uv}[B_0, B_1, B_2, \omega, M]$. It remains to apply Lemma 4.1 again to compute the table $t[]$ that represents the data with respect to the standard ordering. This completes the necessary work for an introduce edge bag.

**Runtime.** Let us now analyze the time required to compute all table entries for one bag. We already know how to compute $t[B_0, B_1, B_2, \omega, \cdot]$ in time $2^{|B_1|/2-1}|B_1|^{\mathcal{O}(1)}$, spending effectively $|B_1|^{\mathcal{O}(1)}$ per table entry (each entry corresponds to one basis matching on $B_1$). For every weight $1 \leq \omega \leq nw_{\max}$ the number of tuples $(B_0, B_1, B_2, M)$ where $B_0, B_1, B_2$ partitions of the bag $B$ and $M$ is one of the $2^{|B_1|/2-1}$ basis matchings equals $(2 + \sqrt{2})^{|B|}$ by the multinomial theorem, and hence we have proved the following:

**Lemma 4.2.** *There is an algorithm that given a graph $G$ along with a path decomposition of width $\mathtt{pw}$ computes the table entries $t[]$ corresponding to the rightmost introduce edge bag of the path decomposition in $(2 + \sqrt{2})^{\mathtt{pw}} w_{\max} (n \cdot \mathtt{pw})^{\mathcal{O}(1)}$ time.*

Now, we can wrap up by completing our algorithm.

**Theorem 4.3.** *There exists a Monte-Carlo algorithm that given a graph $G$ along with a path decomposition of width $\mathtt{pw}$ solves the Hamiltonian cycle problem in $(2+\sqrt{2})^{\mathtt{pw}}(n \cdot \mathtt{pw})^{\mathcal{O}(1)}$ time. The algorithm cannot give false positives and may give false negatives with probability at most $1/2$.*

*Proof.* The algorithm is the following: First guess an edge $\{u,v\}$ in the solution that is incident on the vertex $v$ introduced in the rightmost introduce vertex bag, and reorder the path decomposition as described above. (The guessing is of course implemented by trying all of these at most $\mathtt{pw}$ many edges.) Assign for each edge $e \in E$ a weight $\omega(u) \in \{1, 2, \ldots, 2|E|\}$ uniformly and independently at random. Then use Lemma 4.2 to compute the table entry $t[\emptyset, \{u,v\}, \emptyset, \{\{u,v\}\}, w]$ corresponding to the rightmost introduce edge bag for every weight $w \leq n \cdot w_{\max} \leq n2|E|$ in $2^{t/2}(nt)^{\mathcal{O}(1)}$; Return YES if one of these values is 1 and NO otherwise.

This procedure clearly runs in the claimed running time. If the algorithm return YES, there exists a Hamiltonian cycle including the edge $\{u,v\}$ since there exists a Hamiltonian cycle by the definition of $t$. Otherwise, suppose that there exists a Hamiltonian cycle. Then with probability at least $1 - |E|/2|E| = 1/2$ we have by Lemma 3.9 that $\omega$ isolates the family of all Hamiltonian cycles of $G$. Hence with probability at least $1/2$ there exists a unique minimum weight Hamiltonian cycle $H$ minimizing $\omega(H)$. Let $\{u,v\}$ be the edge incident to $v$ contained in $H$. Then we have that $t[\emptyset, \{u,v\}, \emptyset, \{(u,v)\}, \omega(H) - \omega(\{u,v\})] = 1$ since $H$ is the only Hamiltonian cycle of weight $\omega(H)$. □

### Further results

Combining the ideas from the proof of Theorem 4.3 with Lemma 4.4 and a trick from [15], we will now give an application to fast exponential time algorithms for the Hamiltonian cycle problems on graphs of degree at most 3. To obtain this we require the following result:

**Theorem 4.4** ([19]). *For any $\epsilon > 0$ there exists an integer $n_\epsilon$ such that for any graph $G$ with $n > n_\epsilon$ vertices,*

$$\mathtt{pw}(G) \leq \frac{1}{6}n_3 + \frac{1}{3}n_4 + \frac{13}{30}n_5 + n_{\geq 6} + \epsilon n,$$

*where $n_i$ is the number of vertices of degree $i$ in $G$ for any $i \in \{3, \ldots, 5\}$ and $n_{\geq 6}$ is the number of vertices of degree at least 6.*

This theorem is constructive, and the corresponding path decomposition (and, consequently, tree decomposition) can be found in polynomial time. The observation from [15] is that on cubic graphs, we know that the degree of a vertex in a partial cycle cover of the type used in the our table entries cannot be 2 if at most 1 incident edges is introduced, and it cannot be 0 if at least 2 edges are introduced. Hence these states can be safely ignored since they will be empty. Summarizing, we obtain the following:

**Corollary 4.5.** *There exists a Monte-Carlo algorithm that given a graph $G$ of maximum degree three solves the Hamiltonian cycle problem in $(1+\sqrt{2})^{n/6}n^{\mathcal{O}(1)} \leq 1.1583^n n^{\mathcal{O}(1)}$ time. The algorithm cannot give false positives and may give false negatives with probability at most $1/2$.*

Again the proof idea similar to proof of Theorem 4.3 we can also rather easily obtained the following result:

**Theorem 4.6.** *There exist algorithms that given a graph $G$ and integer $k$ finds a path of length $k$ in time $n(2+\sqrt{2})^{\mathtt{pw}}(k+\mathtt{pw})^{\mathcal{O}(1)}$ time if a path decomposition of width $\mathtt{pw}$ of $G$ is given.*

The idea is to slightly modify the table entries defined in this section: In the definition we considered partial cycle covers visiting all already forgotten vertices. We allow to not visit some vertices and keep track of the number of visited vertices with an additional counter, and hence the formula in the forget bag has to be altered to $t[B_0, B_1, B_2, \omega, M, \ell] \equiv t'[B_0, B_1, B_2 \cup \{v\}, \omega, M, \ell] + t'[B_0 \cup \{v\}, B_1, B_2, \omega, M, \ell]$ where $\ell$ is the additional counter. The additional counted can be bounded by the observation that a non-zero entry with $\ell \geq \mathtt{pw} \cdot k$ directly guarantees a solution.

## 5 Lower bound for Hamiltonian cycle

In this section we prove that the runtime of our algorithm from Section 4 is essentially tight. We show that no algorithm can achieve a significantly better dependence on the pathwidth of the input graph (even at the cost of a larger polynomial factor in the input size), without giving also a breakthrough result for solving CNF-Sat; this is expressed by the main theorem of this section.

**Theorem 5.1.** *If any algorithm solves the Hamiltonian cycle problem in $(2 + \sqrt{2} - \epsilon)^{\mathtt{pw}}|V|^{\mathcal{O}(1)}$ time, then there exists an algorithm that solves the CNF-Sat problem in $(2 - \epsilon)^n m^{\mathcal{O}(1)}$ time.*

Recall the definition of the Matching Connectivity matrix $\mathcal{H}_t$: All rows and columns are indexed by perfect matchings $M_1, M_2$ of the complete graph on $t$ vertices and $\mathcal{H}_t[M_1, M_2] = [M_1 \cup M_2$ is a Hamiltonian cycle]. Our proof, namely a reduction from CNF-Sat, uses the lower bound on the rank of $\mathcal{H}_t$, in particular, the fact that the family of basis matchings $\mathbf{X}_t$ from Section 3 induce a permutation submatrix in $\mathcal{H}_t$ (see Proposition 3.3).

### 5.1 Gadgets

In this section we will introduce three gadgets used for the final construction. Since our gadgets accept some parameters we will call a graph that contains some gadgets an *annotated graph*.

**Vertices with labeled incident edges.** The following two gadgets allow to label incident edges of a vertex $v$ with a labeling function $\lambda_v$, ensuring that every Hamiltonian cycle enters and leaves a vertex with edges of the same label.

**Definition 5.2.** A label gadget is a pair $(v, \lambda_v)$ where $\lambda_v$ is a labeling of the edges incident to $v$. A Hamiltonian cycle $C$ is *consistent* with a label gadget $(v, \lambda_v)$ if $\lambda_v(e) = \lambda_v(e')$ where $e, e'$ are the two edges of $C$ incident to $v$.

The first gadget is for a vertex whose incident edges have only two distinct labels, denoted by dotted and normal lines in figures. It is shown at the left-hand size in Figure 2. We have to ensure that any Hamiltonian cycle contains exactly two edges of the set of edges leaving the gadget and that they are of the same label, and this can be seen to hold by a simple case analysis: If the cycle enters $v_1$, it must continue with $v_2, v_3$. Then it cannot leave the gadget, because then it is impossible to visit all six remaining vertices. Hence it must continue with $v_6$ and then $v_5, v_4, v_9, v_8, v_7$ is forced. The cases where it enters at a different vertex are symmetric.

Let us continue with the multilabel gadget, which we will first formally describe. Given a vertex $v$ in a graph $G$ with incident edges $X$ and a labeling $\lambda_v : X \to [k]$, we obtain $G'$ by replacing $v$ as follows:

- Create a cycle of $3k$ vertices consecutively denoted by $v_1^a, v_1^b, v_1^c, v_2^a, v_2^b, v_2^c, \ldots, v_k^c$.

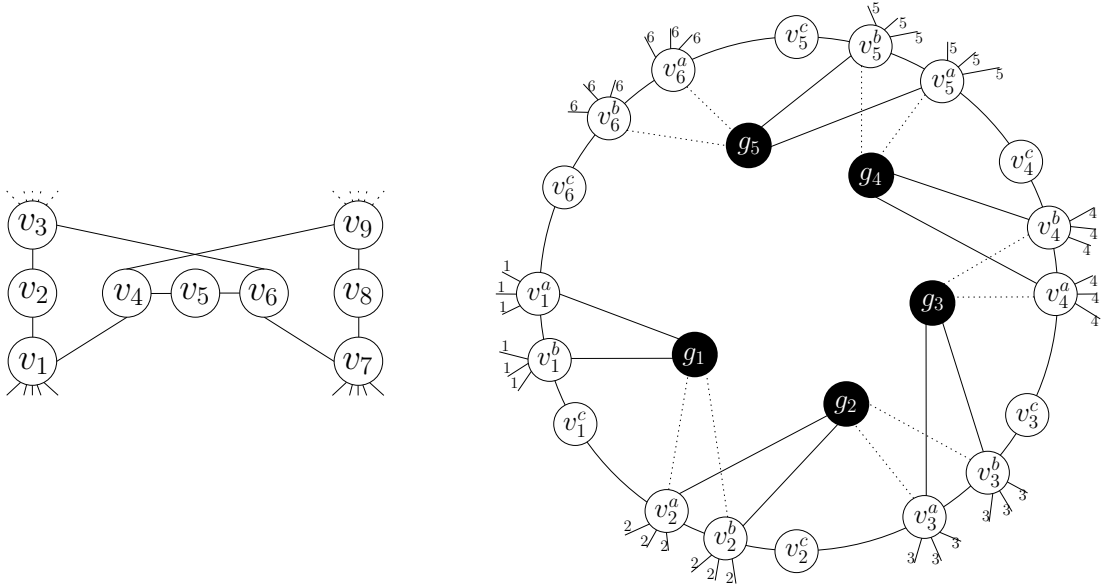

Figure 2: Gadget for vertices with labeled incident edges. On the left-hand side the 2-label gadget is shown. The two label classes are denoted by normal and dotted lines. On the right-hand side the multi-labeled gadget for six labels is shown. The black vertices represent 2-label gadgets and the labels of their incident edges are denoted by normal and dotted lines. A number $i$ next to a leaving edge indicates the edge represents an edge in the original graph with label $i$.

- For every $1 \leq i \leq k$, and every vertex $w \neq v$ incident to an edge $e \in X$ such that $\lambda_v(e) = i$, add the edges $v_i^a w$ and $v_i^b w$.

- For every $1 \leq i < k$ add a 2-label gadget consisting of vertex $g_i$, edges $g_i v_i^a$, $g_i v_i^b$ labeled by 1 and the edges $g_i v_{i+1}^a$, $g_i v_{i+1}^b$ labeled by 2. The $g_i$ are called *guard vertices*.

An instructive example with $k = 6$ is shown on the right-hand side of Figure 2.

**Lemma 5.3.** *There is a Hamiltonian cycle of $G$ consistent with $(v, \lambda_v)$ if and only if $G'$ has a Hamiltonian cycle.*

*Proof.* Note that in a Hamiltonian cycle a guard vertex $g_i$ is either adjacent to both $v_i^a$ and $v_i^b$ or to both $v_{i+1}^a$ and $v_{i+1}^b$, and all the pairs can be adjacent to at most one guard vertex since otherwise there is a cycle on four vertices. Moreover, since each vertex $v_i^c$ is of degree two, any Hamiltonian cycle of $G'$ contains the edges $v_i^b v_i^c$ for $1 \leq i \leq k$, as well as the edges $v_i^c v_{i+1}^a$ for $1 \leq i < k$ and finally the edge $v_k^c v_1^a$. Therefore, since all guard vertices must be visited, any Hamiltonian cycle of $G'$ contains only two edges leaving the gadget and they must be incident to $v_i^a$ and $v_i^b$ for some $i$.

For the forward direction, let there be a Hamiltonian cycle of $G$ that enters and leaves at edges incident of $v$ with label $i$. Then we have to show there is a Hamiltonian path $P$ in the gadget from $v_i^a$ to $v_i^b$, ignoring edges leaving it. Notice that such a $P$ exist in which (i) $g_j$ is adjacent to $v_j^a$ and $v_j^b$ for $j < i$, and (ii) $g_j$ is adjacent to $v_{j+1}^a$ and $v_{j+1}^b$ for $j \geq i$. □

**Induced subgraph gadget.** We will now discuss the induced subgraph gadget, which is shown in Figure 3. The function of the gadget is described as follows:

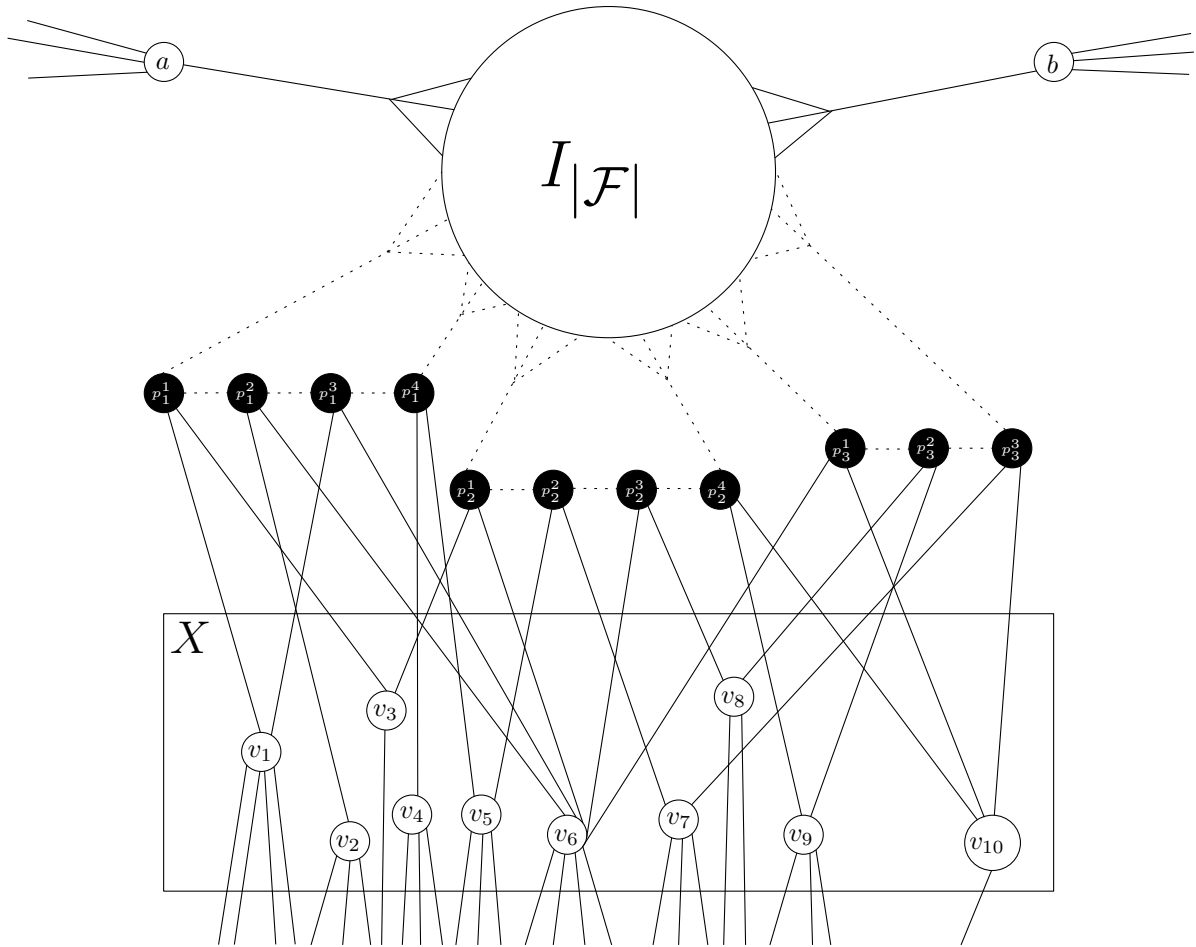

Figure 3: Example of the induced subgraph gadget, with $X = \{v_1, \ldots, v_{10}\}$ and $\mathcal{F} = \{\{v_1v_3, v_2v_6, v_1v_6, v_4v_5\}, \{v_3v_6, v_5v_7, v_6v_8, v_9v_{10}\}, \{v_6v_{10}, v_8v_9, v_7v_{10}\}\}$. The double triangles at the $I_\mathcal{F}$ indicate that in fact there are edges to every vertex in $I_\mathcal{F}$. Label 1 edges are drawn dotted while label 2 edges are drawn normally.

**Definition 5.4.** An induced subgraph gadget in a graph $G = (V, E)$ is described by a tuple $(X, a, b, \mathcal{F})$ where $X \subseteq V$, $a, b \in V \setminus X$, $\mathcal{F} \subseteq 2^{E(X,X)}$ and $\emptyset \notin \mathcal{F}$. A Hamiltonian cycle $C$ of $G$ is *consistent with* $(X, a, b, \mathcal{F})$ if $ab \in C$ and $C \cap E(X, X) \in \mathcal{F}$.

The gadget is implemented with the following construction, obtaining a modified graph $G'$ from $G$:

- remove all edges $E(X, X)$ from the graph $G$,

- add a set $I_{|\mathcal{F}|}$, i.e. an independent set of $|\mathcal{F}|$ vertices to $G$, and make all its vertices adjacent to both $a$ and $b$,

- Let $\mathcal{F} = \{F_1, \ldots, F_\ell\}$, and for $i = 1, \ldots, \ell$ do the following:
    - Let $F_i = \{e_1, \ldots, e_z\}$.
    - Add a path of 2-label gadgets $P_i = \{p_i^1, \ldots, p_i^{|F_i|}\}$, with all edges having label 1.
    - For every vertex $v$ of the independent set $I_{|\mathcal{F}|}$, add an edge with label 1 to $p_i^1$ and $p_i^{|F_i|}$.
    - For $j = 1, \ldots, |F_i|$ add edges $p_i^j x$ and $p_i^j y$ with label 2, where $e_j = xy$.

We will now prove the function of the gadget. The vertices $a$ and $b$ are not useful for using the gadget but are required to be able to implement it.

**Lemma 5.5.** *There exists a Hamiltonian cycle $C$ in $G$ that is consistent with $(X, a, b, \mathcal{F})$ if and only if there exists a Hamiltonian cycle in $G'$.*

The intuition of this is that in a Hamiltonian cycle of the graph the part of the cycle that goes from $a$ to $b$ visits all vertices from the independent set $I_{|\mathcal{F}|}$ and all paths except the one corresponding to corresponding to the element of $\mathcal{F}$ that is chosen.

*Proof.* For the forward direction, let $F_i \in \mathcal{F}$ be the intersection of $C$ with $E(X, X)$. Then the Hamiltonian cycle in $G$ can be extended to a Hamiltonian cycle in $G'$ by replacing every edge $e_j \in F_i$ with two edges $up_i^j$ and $p_i^j v$ where $e_j = uv$ and by replacing the edge $ab$ by the path $a, g_1, P_1, g_2, P_2, \ldots, P_{i-1}, g_i, P_{i+1}, g_{i+1}, \ldots, P_{|\mathcal{F}|} g_{|\mathcal{F}|}, b$, where $g_1, \ldots, g_{|\mathcal{F}|}$ is an arbitrary ordering of $I_{|\mathcal{F}|}$.

For the reverse direction, let $C'$ be a Hamiltonian cycle of $G'$. Observe that since $I_{|\mathcal{F}|}$ is independent, the Hamiltonian cycle $C'$ contains exactly $2|\mathcal{F}|$ edges between $I_{|\mathcal{F}|}$ and $N_{G'}(I_{|\mathcal{F}|}) = \{p_i^1, p_i^{|F_i|} : 1 \leq i \leq |\mathcal{F}|\} \cup \{a, b\}$. By the 2-label gadgets, we know that if for some $1 \leq i \leq |\mathcal{F}|$ the Hamiltonian cycle $C'$ contains an edge between $I_{|\mathcal{F}|}$ and $\{p_i^1, p_i^{|F_i|}\}$, then $C'$ contains exactly one edge between $I_{|\mathcal{F}|}$ and $p_i^1$ as well as exactly one edge between $I_{|\mathcal{F}|}$ and $p_i^{|F_i|}$. However $C'$ cannot contain $2|\mathcal{F}|$ edges between $I_{|\mathcal{F}|}$ and $\{p_i^1, p_i^{|F_i|} : 1 \leq i \leq |\mathcal{F}|\}$, because then it would not visit any other vertices. Furthermore $C'$ cannot contain two edges between $a$ (or $b$) and $I_{|\mathcal{F}|}$, since then it would have to contain either zero or two edges between $b$ (or $a$) and $I_{|\mathcal{F}|}$, making it impossible to visit the original vertices of $G$. Consequently $C'$ contains exactly one edge between $a$ and $I_{|\mathcal{F}|}$, exactly one edge between $b$ and $I_{|\mathcal{F}|}$, and exactly two edges between $P_i$ and $I_{|\mathcal{F}|}$ for $|\mathcal{F}| - 1$ indices $i \in [|\mathcal{F}|]$. So there is exactly one index $1 \leq i_0 \leq |\mathcal{F}|$ such that $p_i^1$ and $p_i^{|F_i|}$ are not connected with an edge of $I_{|\mathcal{F}|}$. Since $C'$ visits all the vertices of $P_{i_0}$ by label 2 edges only, we can obtain a Hamiltonian cycle $C$ in $G$ by removing all the edges of the gadget and adding the edge $ab$ together with edges of $F_{i_0}$. Note that the Hamiltonian cycle $C$ is consistent with $(X, a, b, \mathcal{F})$ as it does not contain any edge of $E(X, X) \setminus F_{i_0}$ since the graph $G'$ has the edges $E(X, X)$ removed. □

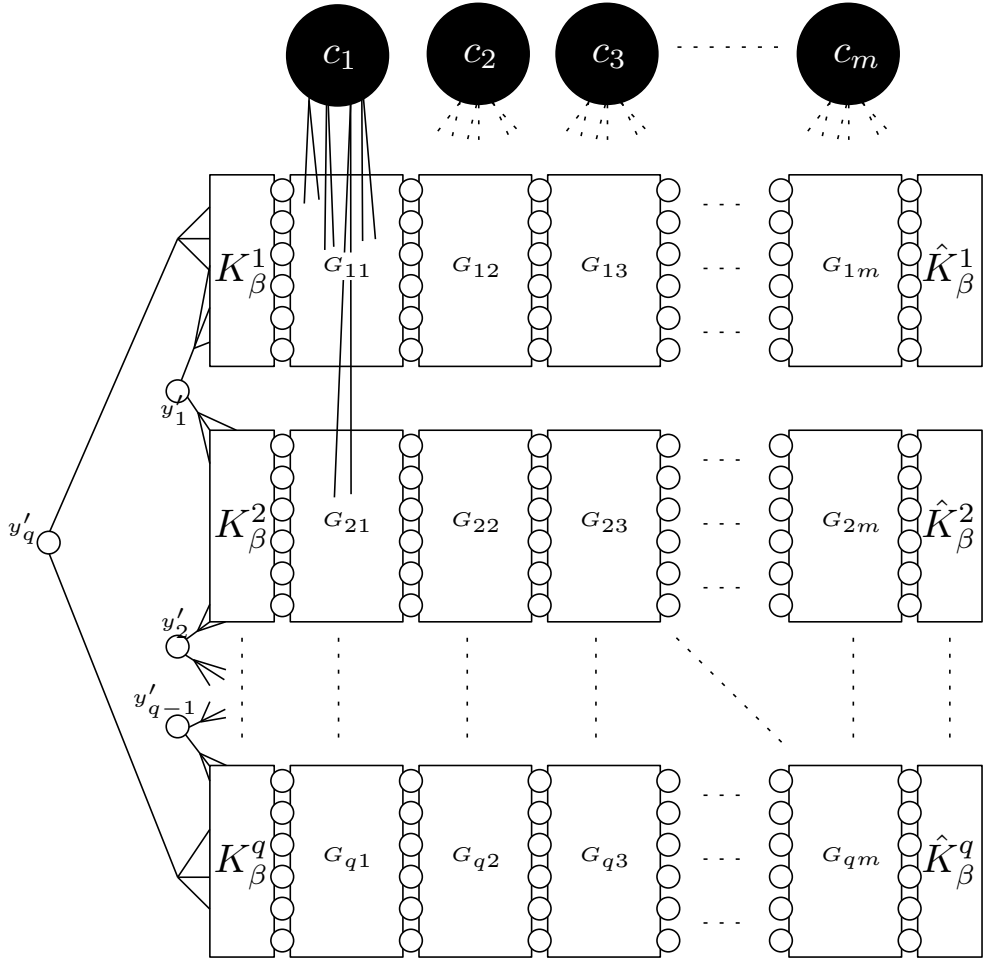

Figure 4: The total construction for the Hamiltonian path lower bound. The $K_\beta$ denote cliques of $\beta$ vertices. The black circles $c_i$ are multilabel gadgets (see Figure 2) and the rectangles $G_{ij}$ are induced subgraph gadgets, used as further illustrated in Figure 5.

### 5.2 Construction

We will now describe the construction of the reduction, see also Figure 4. After the formal definition we will provide some intuition. Our construction is parameterized by two integer constants $\beta$ and $\gamma$, which will be chosen later, such that

$$\sum_{\substack{i_0+i_1+i_2=\beta \\ i_1 \text{ is even}}} \binom{\beta}{i_0, i_1, i_2} \sqrt{2}^{i_1} \geq 2^\gamma. \tag{1}$$

Assume we are given a CNF-formula $\phi = C_1 \wedge \ldots \wedge C_m$ on variables $x_1, \ldots, x_n$ with $n$ being a multiple of $\gamma$; let $q = \frac{n}{\gamma}$. Partition the set $\{x_1, \ldots, x_n\}$ into $n/\gamma$ blocks of size $\gamma$, denoted $X_1, \ldots, X_{n/\gamma}$. Also, denote $X_i = \{x_{i1}, \ldots, x_{i\gamma}\}$. Intuitively, we will represent the $2^\gamma$ assignments of a block of variables by the states of group of vertices in a bag of the to be constructed path decomposition. In Subsection 5.5, we will discuss how to fix $\beta$ and $\gamma$ such that the number of these states, represented on the left hand side of Equation (1) is at least $2^\gamma$.

In the following we will use the family of matchings $\mathbf{X}_t$ defined in Definition 3.1. For ease of notation we denote $\mathbf{X}(S)$ for the family obtained from $\mathbf{X}_{|S|}$ by replacing the elements of $U_t$ one to one with elements of $S$ (using an arbitrary but fixed ordering), for any set $S$ of even cardinality.

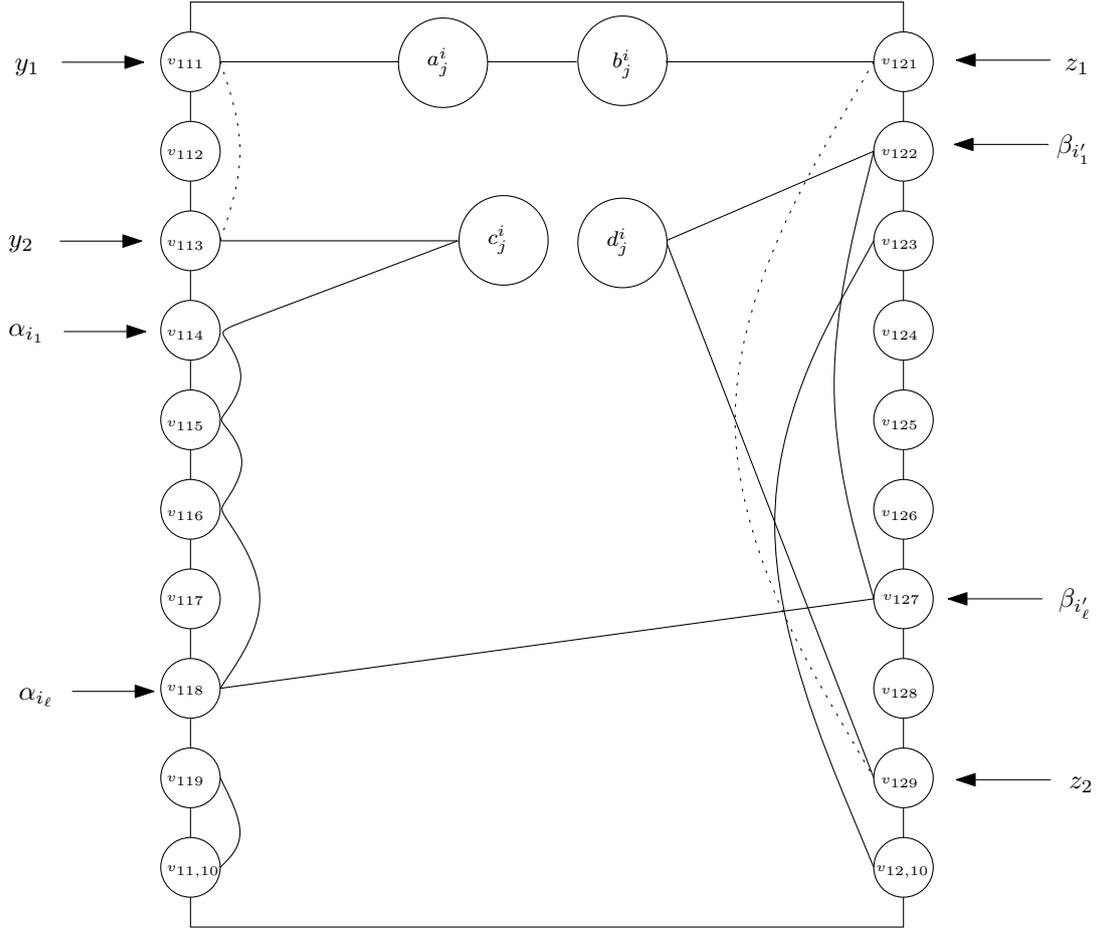

Figure 5: An edgeset from $\mathcal{F}$ in an induced subgraph gadget $G_{11}$ is shown here. For convenience, $v_{1,10}$ and $v_{2,10}$ are denoted by $v_{10}$ and $v_{20}$. The edgeset shown is constructed from the coloring $s$ with $s(v_{112}) = s(v_{117}) = 0$, $s(v_{114}) = s(v_{115}) = s(v_{116}) = s(v_{118}) = 2$ and the matching $(v_{111}, v_{113}), (v_{119}, v_{11,10})$.

1. For every $i = 1, \ldots, q$, $j = 1, \ldots, m+1$ and $k = 1, \ldots, \beta$ add vertices $v_{i,j,k}$; for $i = 1, \ldots, q$ and $j = 1, \ldots, m+1$ denote $V_{ij} = \{v_{ij1}, \ldots, v_{ij\beta}\}$.

2. For every $i = 1, \ldots, q$ add a clique on $\beta$ vertices $K_\beta^i$, and add an edge between all vertices of the clique $K_\beta^i$ and all vertices from $V_{i,1}$.

3. For every $i = 1, \ldots, q$ add a clique on $\beta$ vertices $\hat{K}_\beta^i$, and add an edge between all vertices of the clique $\hat{K}_\beta^i$ and all vertices from $V_{i,m+1}$.

4. For every $i = 1, \ldots, q-1$ add a vertex $y_i'$ and make it adjacent to all vertices of $K_\beta^i$ and $K_\beta^{i+1}$. Furthermore, a vertex $y_q'$ is added and made adjacent all vertices of $K_\beta^q$ and $K_\beta^1$.

5. For every $i = 1, \ldots, q$ and $j = 1, \ldots, m$ we use an induced subgraph gadget $G_{ij}$ (the special vertices $a_j^i$ and $b_j^i$ from Definition 5.4 are included in $G_{ij}$) as follows (see also Figure 5):

(a) Add four vertices $a_j^i, b_j^i, c_j^i, d_j^i$ and the edge $(a_j^i, b_j^i)$, where the edge $(a_j^i, b_j^i)$ is guaranteed to be in any Hamiltonian cycle (by for example subdividing it) as required by the gadget.

(b) The set $X$ consists of the vertices $V_{ij} \cup V_{i,j+1}$ and the vertices $c_j^i, d_j^i$.

(c) Given a vector $s \in \{0,1,2\}^\beta$ and a matching $M \in \mathbf{X}(s^{-1}(1))$ we denote $s_{\langle i,j \rangle}$ as the coloring of $V_{ij}$ and $M_{\langle i,j \rangle}$ as the matching of $V_{ij}$ obtained by identifying the sets $\{1, \ldots, \beta\}$ and $V_{ij}$. (Equivalently, we may take $s_{\langle i,j \rangle} \in \{0,1,2\}^{V_{ij}}$ and $M_{\langle i,j \rangle} \in \mathbf{X}(s_{\langle i,j \rangle}^{-1}(1))$.)

(d) For a every vector $s \in \{0,1,2\}^\beta$ and a matching $M \in \mathbf{X}(s^{-1}(1))$, add an edgeset $\eta_{ij}(s, M)$ constructed as follows to the family $\mathcal{F}$ used for the induced subgraph gadget:

   i. add the edges of the path $\alpha_1, \ldots, \alpha_\ell$ where $\{\alpha_1, \ldots, \alpha_\ell\}$ are the elements of $s_{\langle i,j \rangle}^{-1}(2)$ sorted ascendantly (the ordering is in fact immaterial so long as we have a simple path through these vertices);
   
   ii. add the edges of the path $\beta_1, \ldots, \beta_{\ell'}$ where $\{\beta_1, \ldots, \beta_{\ell'}\}$ are the elements of $s_{\langle i,j+1 \rangle}^{-1}(0)$ sorted ascendantly;
   
   iii. let $e = (y_1, y_2)$ be the edge of $M_{\langle ij \rangle}$ incident with the smallest element of $s_{\langle i,j \rangle}^{-1}(1)$;
   
   iv. add all edges of $M_{\langle ij \rangle}$ not equal to $e$;
   
   v. let $M' \in \mathbf{X}(s^{-1}(1))$ be the unique matching such that $M \cup M'$ is a Hamiltonian cycle (this exists and is unique due to Proposition 3.3);
   
   vi. let $e' = (z_1, z_2)$ be the edge of $M'_{\langle i,j+1 \rangle}$ incident with the smallest element of $s_{\langle i,j \rangle}^{-1}(1)$;
   
   vii. add all edges of $M'_{\langle i,j+1 \rangle}$ not equal to $e'$;
   
   viii. add the edges $(y_1, a_j^i), (y_2, c_j^i), (c_j^i, \alpha_1); (z_1, b_j^i), (z_2, d_j^i), (d_j^i, \beta_1); (\alpha_\ell, \beta_{\ell'})$.

6. Fix arbitrarily an injective mapping
$$\psi : \{0,1\}^\gamma \to \left\{ (\mathbf{s}, M) \,\middle|\, s \in \{0,1,2\}^\beta \wedge |s^{-1}(1)| \text{ is even} \wedge M \in \mathbf{X}(s^{-1}(1)) \right\}.$$

7. For $j = 1, \ldots, m$:

   (a) Add a multilabel gadget $c_j$.
   
   (b) For every $i = 1, \ldots, q$ and every partial assignment $x \in \{0,1\}^{X_i}$ of the variables $X_i$ that satisfies $C_j$, consider the first vertex $\rho$ of the path in the induced subgraph gadget $G_{ij}$ corresponding with the element $\eta_{ij}(\psi(x))$ of $\mathcal{F}$.
   
   (c) Let $(\rho, \sigma)$ be an edge with label 1 in the construction of the induced subgraph gadget. Add an edge $(\rho, c_j)$ and $(c_j, \sigma)$ such that with respect to $\rho$ and $\sigma$ the edges again have label 1 and with respect to the multilabel gadget $c_j$ the edges have a label that has not been used yet in the interval $[j \cdot 2^\gamma + 1, (j+1) \cdot 2^\gamma]$, which must exist since there are at most this many assignments of $X_i$.

The intuition behind the construction is as follows: $V_{ij}$ are blocks of vertices whose joint states encode a joint assignment of $X_i$. In Step $2 - 4$ vertices are added to ensure that a set of disjoint paths visiting all vertices in the remainder of the graph can be completed into a Hamiltonian cycle. In Step 5 we create graphs $G_{ij}$ on $V_{ij}$ and $V_{i,j+1}$ such that any for any set of these disjoint paths, the intersection of $G_{ij}$ and $G_{i,j+1}$ is similar. To do this we use the induced subgraph gadget and for every state $(s, M)$ we allow exactly one edgeset in $\mathcal{F}$ that induces state $(s, M)$. In Step 6 we fix an encoding $\psi$ of partial assignments in states of the group of vertices $V_{ij}$. As mentioned before, this encoding exists due to (1). Finally, we check in Step 7 whether clause $j$ is satisfied by adding a multilabel gadget $c_j$ that can be visited if and only if in some graph $G_{ij}$ the intersection $X$ from the family $\mathcal{F}$ induces a state $(c, M) = \eta_{ij}^{-1}(X)$ such that the partial assignment $\psi^{-1}(s, M)$ satisfies clause $C_j$. The labels to the clause gadget are chosen consecutively in 7c to ensure adding the labels only increases the pathwidth by a constant.

## 5.3 From a satisfying assignment to a Hamiltonian cycle

Suppose that $x \in \{0, 1\}^n$ satisfies the formula $\phi$. Then we first claim that for every $i = 1, \ldots, q$ the set

$$\mathcal{E} = \bigcup_{j=1}^{m} \eta_{ij}(\psi(X_i)) \qquad (2)$$

is a set of disjoint paths in the constructed graph $G$, in which all vertices of $\cup_{j=1}^{m} G_{ij}$ are visited and all endpoints are in $V_{i,1} \cup V_{i,m+1}$. To this end we first need to zoom in at the application of the induced subgraph gadget in Step 6. of the construction (see also Figure 5). Note that effectively, in a set $\eta_{i,j}(s, M)$, the first vertex $y_1$ of $s_{\langle ij \rangle}(-1)$ is connected to the first vertex $z_1$ of $s_{\langle i,j+1 \rangle}(-1)$ through $a_j^i, b_j^i$. Recall that $y_2$ is the second vertex of the lexicographically first edge of $M_{\langle i,j \rangle}$. Suppose that $y_2 = v_{ijk}$.

Then, to show the claim on $\mathcal{E}$, it is sufficient to show that the second vertex $y_2$ is connected to $v_{i+1,j}$ by a path with internal vertices $s_{\langle ij \rangle}^{-1}(2) \cup s_{\langle i,j+1 \rangle}^{-1}(\{0,1\}) \cup c_j^i, d_j^i$. This follows by construction: in $V_{i+1}$ we use two matchings on the vertices with label one that are guaranteed to form a Hamiltonian cycle so since we left out the edges $(z_1, z_2)$ in one matching and the edge $(z_1, v_{i,j+1,k})$ they must now form a path between the $z_2$ and $v_{i,j+1,k}$. It is easily seen that the further extension of the path meets the claim.

Now we alter the set $\mathcal{E}$ to a set $\mathcal{E}'$ to visit all clause multilabel gadgets as follows: since $x$ satisfies the formula, we know by construction that for every $j = 1, \ldots, m$ there exists $i = 1, \ldots, q$ such a variable of $X_i$ is satisfying $C_j$. We take one such $i$ and we know that in the path created in the induced subgraph gadget corresponding to the element $\eta(\psi(X_{ij}))$ of $\mathcal{F}$ we added in Step 7.C the two edges to the clauses gadget with the same label between a vertex $\rho$ of $X$ used and the first vertex $\sigma$ of the created path. Since we know $(\rho, \sigma) \in \mathcal{E}$ we can safely replace $(\rho, \sigma)$ with $(\rho, c_j)$ $(c_j, \sigma)$.

Hence now $\mathcal{E}'$ is a set of disjoint paths, with endpoints in $\cup_{i=1}^{q} V_{i,1} \cup V_{i,m+1}$, all vertices $c_1, \ldots, c_m$ and vertices from $\cup_{i,j} G_{ij}$ are visited, and by construction it is consistent with all the multilabel and induced subgraph gadgets. After contracting the internal points of these paths, it is easy to see we are left with a graph that has a Hamiltonian cycle: for every $i = 1, \ldots, q$ we can use the cliques to extend the paths from $\mathcal{E}'$ to a path between two vertices of $K_\beta^i$ that contains all paths from $\mathcal{E}'$ with endpoints in $\cup_{i=1}^{q} V_{i,1} \cup V_{i,m+1}$ that visits $K_\beta^i \cup \hat{K}_\beta^i$ and does not visit $y'_{i-1}$ and $y'_i$ such that the vertices $y'_i$ then can be used to connect these paths into a complete Hamiltonian cycle. By construction it is easily seen to be consistent with all gadgets.

## 5.4 From a Hamiltonian cycle to a satisfying assignment

Suppose we are given a Hamiltonian cycle $C$ of $G$, consistent with all the induced subgraph and multilabel gadgets.

Contract edges to clause gadgets obtaining $C'$. By the multilabel gadget and the labeling, the edges resulting from the contraction will be contained in an induced subgraph gadget. Let us denote $X_{ij}$ for the vertex set of $G_{ij}$ minus $a_{ij}, b_{ij}$.

**Lemma 5.6.** *For every $i$ and $j$ it holds that*

$$\eta_{ij}^{-1}(C' \cap (X_{ij} \times X_{ij})) = \eta_{i,j+1}^{-1}(C' \cap (X_{i,j+1} \times X_{i,j+1})).$$

*Proof.* Let $(\mathbf{s}, M) = \eta_{ij}^{-1}(C' \cap (X_{ij} \times X_{ij}))$ and $(\mathbf{s}', M') = \eta_{i,j+1}^{-1}(C' \cap (X_{i,j+1} \times X_{i,j+1}))$. First, we know that for every $s_i + s'_i = 2$ since otherwise the vertices of $V_i$ will not have degree 2 in $C'$. Hence, we know that $s^{-1}(1) = (s')^{-1}(1)$. If $M \neq M'$ then we know by Proposition 3.3 that $M \cup M'$ is not a Hamiltonian cycle since the matchings are by construction from $\mathbf{X}(s^{-1}(1))$. Hence we have that $M \cup M'$ contains two subcycles and the one avoiding the element 1 will give a subcycle of $C'$, contradicting that $C$ is a Hamiltonian cycle. □

Hence, the cycle $C'$ must represent an assignment of the variables $x_1, \ldots, x_n$. Furthermore, since $C$ visits all multilabel gadgets, it must represent a satisfying assignment by construction.

## 5.5 Pathwidth and efficiency bound

We will first argue that the pathwidth of the constructed graph (after expanding all gadgets) $G$ is $\left\lceil \frac{n}{\gamma} \right\rceil \beta + f(\beta, \gamma)$ for some function $f$ and then set the parameters $\beta$ and $\gamma$ in order to prove the theorem.

Let $\kappa = 1, \ldots, q$ and $j = 1, \ldots, m$. In the multilabel gadget $c_j$, recall there is a vertex $v_1^a$, adjacent to some vertex in the induced subgraph gadget $G_{1j}$. Similarly, let $\ell$ be the maximum integer such that in this multilabel gadget, $v_\ell^a$ is adjacent to a vertex in one of the induced subgraph gadgets $G_{1j}, \ldots, G_{\kappa,j}$. Then note that

$$S_{\kappa,j} = (\bigcup_{i=1}^{\kappa} V_{ij}) \cup (\bigcup_{\kappa=i+1}^{q} V_{i,j+1}) \cup \{v_1^a, g_\ell, v_\ell^c\},$$

is a separator since the labels in the multilabel gadget are ordered such that labels coming from the induced subgraph gadget are adjacent. Then the claimed upper bound on the pathwidth follows easily by using the above separators as bags consecutively for every $j = 1, \ldots, m$ and $\kappa = 1, \ldots, q$ since the size of every union of two consecutive separators is $\left\lceil \frac{n}{\gamma} \right\rceil \beta + f(\beta, \gamma)$ and the components to the left of $S_{1,1}$ and to the right of $S_{q,m}$ are of constant pathwidth.

Now we will discuss how to fix the constants $\beta$ and $\gamma$. Note that by splitting summations we obtain

$$\sum_{\substack{i_0+i_1+i_2=\beta \\ i_1 \text{ is even}}} \binom{\beta}{i_0, i_1, i_2} \sqrt{2}^{i_1} = \sum_{\substack{0 \leq i_1 \leq \beta \\ i_1 \text{ even}}} \sqrt{2}^{i_1} 2^{\beta - i_1},$$

and since for every odd $i_1$, we have $\sqrt{2}^{i_1} 2^{\beta - i_1} \leq \sqrt{2}^{i_1} 2^{\beta - i_1 - 1}$ we know by the multinomial theorem that

$$\sum_{\substack{i_0+i_1+i_2=\beta \\ i_1 \text{ is even}}} \binom{\beta}{i_0, i_1, i_2} \sqrt{2}^{i_1} \geq \frac{1}{2} \sum_{i_0+i_1+i_2=\beta} \binom{\beta}{i_0, i_1, i_2} \sqrt{2}^{i_1} = (2 + \sqrt{2})^{\beta - 1}.$$

Hence, we satisfy Equation 1 by setting $\beta = \frac{\gamma}{\lg(2+\sqrt{2})} + 1$. Running the assumed algorithm that solves the Hamiltonian cycle problem in $(2 + \sqrt{2} - \epsilon)^{\mathtt{pw}}|V|^{\mathcal{O}(1)}$ algorithm then takes time

$$(2 + \sqrt{2} - \epsilon)^{\left\lceil \frac{n}{\gamma} \right\rceil \left( \frac{\gamma}{\lg(2+\sqrt{2})} + 1 \right)} \leq 2^{\lg(2+\sqrt{2}-\epsilon) \frac{n+1}{\gamma} \left( \frac{\gamma}{\lg(2+\sqrt{2})} + 1 \right)} = 2^{(n+1) \left( \frac{\lg(2+\sqrt{2}-\epsilon)}{\lg(2+\sqrt{2})} + \frac{\lg(2+\sqrt{2}-\epsilon)}{\gamma} \right)}$$

So choosing $\gamma$ large enough such that $\left( \frac{\lg(2+\sqrt{2}-\epsilon)}{\lg(2+\sqrt{2})} + \frac{\lg(2+\sqrt{2}-\epsilon)}{\gamma} \right) < 1$ is sufficient and gives a $(2 - \epsilon')^n m^{\mathcal{O}(1)}$ algorithm for CNF-Sat, concluding the proof of Theorem 5.1.

# 6 Conclusions

We have presented a set of matchings $\mathbf{X}_n$, which forms a basis of the Matching Connectivity matrix $\mathcal{H}_n$. In particular we have obtained a factorization theorem (Theorem 3.4) which shows an explicit way of expressing any perfect matching as a linear combination of matchings of $\mathbf{X}_n$. As a consequence we obtained deterministic algorithms for computing the parity of the number of Hamiltonian cycles in undirected graphs and directed bipartite graphs in $\mathcal{O}(1.888^n)$, which together with the Isolation Lemma lead to Monte Carlo algorithms solving the decision versions of Hamiltonicity within the same running time. Moreover, using the basis $\mathbf{X}_n$, we presented an algorithm which given an undirected graph on $n$ vertices along with a path decomposition of width at most $\mathtt{pw}$, decides Hamiltonicity in $(2+\sqrt{2})^{\mathtt{pw}} n^{\mathcal{O}(1)}$ time. Somewhat surprisingly we use the same tool, i.e. the basis $\mathbf{X}_n$, to show by an involved reduction from CNF-Sat that our bounded pathwidth algorithm is optimal under the Strong Exponential Time Hypothesis.

Our results lead to several natural open problems. Can the basis $\mathbf{X}_n$ be used to obtain a deterministic $\mathcal{O}((2-\epsilon)^n)$ time algorithm for Hamiltonicity? Can we handle directed graphs without the bipartiteness assumption? Can we extend our bounded pathwidth algorithm to a bounded treewidth algorithm with the same complexity?

Finally we would like to note that the row space of the Matching Connectivity matrix $\mathcal{H}_n$ clearly has several different bases. We have investigated a particular one, which proved to have several interesting properties and applications, however there might be different ones which are also worth exploring.

# Acknowledgements

We would like to thank Marcin Pilipczuk, Michał Pilipczuk, Johan van Rooij and Jakub Onufry Wojtaszczyk for early discussions on Section 5.

# References

[1] Sanjeev Arora. Polynomial time approximation schemes for Euclidean traveling salesman and other geometric problems. *J. ACM*, 45(5):753–782, September 1998.

[2] Sanjeev Arora and Boaz Barak. *Computational complexity: a modern approach*. Citeseer, 2009.

[3] Eric T. Bax. Inclusion and exclusion algorithm for the Hamiltonian path problem. *Information Processing Letters*, 47(4):203 – 207, 1993.

[4] R.E. Bellman. Combinatorial processes and dynamic programming. 1958.

[5] Richard Bellman. Dynamic programming treatment of the travelling salesman problem. *J. ACM*, 9(1):61–63, 1962.

[6] Andreas Björklund. Determinant sums for undirected Hamiltonicity. In *FOCS*, pages 173–182. IEEE Computer Society, 2010.

[7] Andreas Björklund, Thore Husfeldt, Petteri Kaski, and Mikko Koivisto. Trimmed Möbius inversion and graphs of bounded degree. *Theory Comput. Syst.*, 47(3):637–654, 2010.

[8] Hans Bodlaender. Treewidth: Structure and algorithms. In Giuseppe Prencipe and Shmuel Zaks, editors, *Structural Information and Communication Complexity*, volume 4474 of *Lecture Notes in Computer Science*, pages 11–25. Springer Berlin / Heidelberg, 2007.

[9] Hans L. Bodlaender, Marek Cygan, Stefan Kratsch, and Jesper Nederlof. Solving weighted and counting variants of connectivity problems parameterized by treewidth deterministically in single exponential time. 2012.

[10] Hans L. Bodlaender and Arie M. C. A. Koster. Combinatorial Optimization on Graphs of Bounded Treewidth. *The Computer Journal*, 51(3):255–269, 2008.

[11] Hajo Broersma, Fedor V. Fomin, Pim van 't Hof, and Daniël Paulusma. Fast exact algorithms for Hamiltonicity in claw-free graphs. In Christophe Paul and Michel Habib, editors, *WG*, volume 5911 of *Lecture Notes in Computer Science*, pages 44–53, 2009.

[12] Nicos Christofides. Worst-case analysis of a new heuristic for the travelling salesman problem. Technical Report 388, Graduate School of Industrial Administration, Carnegie Mellon University, 1976.

[13] Bruno Courcelle. The monadic second-order logic of graphs. i. recognizable sets of finite graphs. *Inf. Comput.*, 85(1):12–75, 1990.

[14] Marek Cygan, Holger Dell, Daniel Lokshtanov, Dániel Marx, Jesper Nederlof, Yoshio Okamoto, Ramamohan Paturi, Saket Saurabh, and Magnus Wahlström. On problems as hard as CNF-SAT. In *IEEE Conference on Computational Complexity*, pages 74–84. IEEE, 2012.

[15] Marek Cygan, Jesper Nederlof, Marcin Pilipczuk, Michal Pilipczuk, Johan M. M. van Rooij, and Jakub Onufry Wojtaszczyk. Solving connectivity problems parameterized by treewidth in single exponential time. In Rafail Ostrovsky, editor, *FOCS*, pages 150–159. IEEE, 2011.

[16] Frederic Dorn, Fedor V. Fomin, and Dimitrios M. Thilikos. Catalan structures and dynamic programming in $H$-minor-free graphs. pages 631–640, 2008.

[17] Stuart E. Dreyfus and Robert A. Wagner. The Steiner problem in graphs. *Networks*, 1:195–207, 1972.

[18] David Eppstein. The traveling salesman problem for cubic graphs. 11(1):61–81, 2007.

[19] Fedor V. Fomin, Serge Gaspers, Saket Saurabh, and Alexey A. Stepanov. On two techniques of combining branching and treewidth. 54(2):181–207, 2009.

[20] Fedor V. Fomin and Dieter Kratsch. *Exact Exponential Algorithms*. Springer-Verlag New York, Inc., New York, NY, USA, 1st edition, 2010.

[21] Heidi Gebauer. On the number of Hamilton cycles in bounded degree graphs. pages 241–248, 2008.

[22] Mika Göös and Jukka Suomela. Locally checkable proofs. In Cyril Gavoille and Pierre Fraigniaud, editors, *PODC*, pages 159–168. ACM, 2011.

[23] Michael Held and Richard M. Karp. A dynamic programming approach to sequencing problems. In *Proceedings of the 1961 16th ACM national meeting*, ACM '61, pages 71.201–71.204, New York, NY, USA, 1961. ACM.

[24] Illya V. Hicks, Arie M. C. A. Koster, and Elif Kolotoğlu. Branch and tree decomposition techniques for discrete optimization. *Tutorials in Operations Research 2005*, pages 1–19, 2005.

[25] Russell Impagliazzo, Ramamohan Paturi, and Francis Zane. Which problems have strongly exponential complexity? *J. Comput. Syst. Sci.*, 63(4):512–530, 2001.

[26] Kazuo Iwama and Takuya Nakashima. An improved exact algorithm for cubic graph TSP. pages 108–117, 2007.

[27] Richard M. Karp. Dynamic programming meets the principle of inclusion and exclusion. *Oper. Res. Lett.*, 1:49–51, 1982.

[28] Jon Kleinberg and Éva Tardos. *Algorithm Design*. 2005.

[29] Ton Kloks. *Treewidth, Computations and Approximations*, volume 842 of *Lecture Notes in Computer Science*. Springer, 1994.

[30] Samuel Kohn, Allan Gottlieb, and Meryle Kohn. A generating function approach to the traveling salesman problem. In *ACM '77: Proceedings of the 1977 annual conference*, pages 294–300, New York, NY, USA, 1977. ACM.

[31] S. Lin and B. W. Kernighan. An effective heuristic algorithm for the traveling-salesman problem. *Operations Research*, 21(2):pp. 498–516, 1973.

[32] Daniel Lokshtanov, Daniel Marx, and Saket Saurabh. Known algorithms on graphs of bounded treewidth are probably optimal. pages 777–789, 2011.

[33] Ketan Mulmuley, Umesh V. Vazirani, and Vijay V. Vazirani. Matching is as easy as matrix inversion. 7(1):105–113, 1987.

[34] Rolf Niedermeier. *Invitation to Fixed-Parameter Algorithms*. 2002.

[35] Mihai Pătrașcu and Ryan Williams. On the possibility of faster SAT algorithms. In *Proc. 21st ACM/SIAM Symposium on Discrete Algorithms (SODA)*, 2010. To appear.

[36] Ran Raz and Boris Spieker. On the "log rank"-conjecture in communication complexity. *Combinatorica*, 15(4):567–588, 1995.

[37] Neil Robertson and Paul D. Seymour. Graph minors. III. Planar tree-width. *J. Comb. Theory*, 36(1):49–64, 1984.

[38] Gerhard J. Woeginger. Exact algorithms for NP-hard problems: A survey. In Michael Jünger, Gerhard Reinelt, and Giovanni Rinaldi, editors, *Combinatorial Optimization*, volume 2570 of *Lecture Notes in Computer Science*, pages 185–208. Springer, 2001.

# A  Omitted proofs of Section 3

## A.1  Proof of Theorem 3.4

This section provides a proof for Theorem 3.4, which itself implies that each family $\mathbf{X}_t$ gives a basis for the Matching Connectivity matrix $\mathcal{H}_t$. The proof will be by induction. As a first tool we define a projection that maps *any* perfect matching of $U_t$ to a certain perfect matching of $U_{t-2}$; this is controlled by an additional parameter $b \in \{0, 1\}$. For matchings $X(t, \cdot)$, depending on $b$, this will either be equivalent to undoing the last step in the recursive definition of $X(t, \cdot)$ or yield $\emptyset$ (see Proposition A.2).

**Definition A.1** (Projection to $t-2$ first vertices). Let $t \geq 4$ be an even integer. We define a function
$$\mathtt{shrink}_t \colon (\Pi_2(U_t) \cup \{\emptyset\}) \times \{0, 1\} \to \Pi_2(U_{t-2}) \cup \{\emptyset\}$$
as follows. For $b \in \{0, 1\}$ we let $\mathtt{shrink}_t(\emptyset, b) := \emptyset$. For $M \neq \emptyset$, the definition is as follows; we let $\alpha(i) := \alpha_M(i)$:

1. $\{t-1, t-2\} \in M$:
$$\mathtt{shrink}_t(M, 1) := \emptyset$$
$$\mathtt{shrink}_t(M, 0) := M \setminus \{\{t-1, t-2\}\}$$

2. $\{t-1, t-3\} \in M$:
$$\mathtt{shrink}_t(M, 1) := (M \setminus \{\{t-1, t-3\}, \{t-2, \alpha(t-2)\}\}) \cup \{\{t-3, \alpha(t-2)\}\}$$
$$\mathtt{shrink}_t(M, 0) := \emptyset$$

3. $\{t-2, t-3\} \in M$:
$$\mathtt{shrink}_t(M, b) := (M \setminus \{\{t-1, \alpha(t-1)\}, \{t-2, t-3\}\}) \cup \{\{t-3, \alpha(t-1)\}\}$$

4. $\{t-1, t-2\}, \{t-1, t-3\}, \{t-2, t-3\} \notin M$:
$$\mathtt{shrink}_t(M, 1) := (M \setminus \{\{t-1, \alpha(t-1)\}, \{t-2, \alpha(t-2)\}\}) \cup \{\{\alpha(t-1), \alpha(t-2)\}\}$$
$$\mathtt{shrink}_t(M, 0) := (M \setminus \{\{t-1, \alpha(t-1)\}, \{t-2, \alpha(t-2)\}, \{t-3, \alpha(t-3)\}\}) \cup$$
$$\cup \{\{t-3, \alpha(t-2)\}, \{\alpha(t-1), \alpha(t-3)\}\}$$

We omit the subscript $t$ when it is clear from context.

**Proposition A.2.** *The behavior of $\mathtt{shrink}_t$ for matchings $X(t, ab)$ is a follows*
$$\mathtt{shrink}_t(X(t, ab), c) = \begin{cases} X(t-2, a) & \text{if } b = \bar{c}, \\ \emptyset & \text{else.} \end{cases}$$

Now we show how our $\mathtt{shrink}_t$ operation interacts with the fact whether or not two matchings of $U_t$ form a Hamiltonian cycle. Recall the $\sqcap$-operator: We have that the union of two perfect matchings $M$ and $M'$ of $U_t$ is a Hamiltonian cycle if and only if $M \sqcap M' = \{U_t\}$, i.e., if the connectivity provided by the edges of the two matchings creates a single connected component that contains all vertices $U_t$. (We do not make further use of the meet operator but we find it convenient for specifying also the set of vertices on which we have a Hamiltonian cycle.)

The main fact about this interaction is given by the following lemma. It shows that the fact whether perfect matchings $M_1$ and $M_2$ form a Hamiltonian cycle depends directly on two pairs of projections of the two matchings. Note that the main statement of the lemma is given using addition modulo two.

**Lemma A.3.** *Let $t \geq 4$ be an even integer and let $M_1, M_2 \in \Pi_2(U_t)$. Let $M_i^j := \mathtt{shrink}(M_i, j)$. Then*

$$[M_1 \sqcap M_2 = \{U_t\}] \equiv [M_1^0 \sqcap M_2^1 = \{U_{t-2}\}] + [M_1^1 \sqcap M_2^0 = \{U_{t-2}\}].$$

*In other words, $M_1 \cup M_2$ is a Hamiltonian cycle on $U_t$ if and only if exactly one of $M_1^0 \cup M_2^1$ and $M_1^1 \cup M_2^0$ is a Hamiltonian cycle on $U_{t-2}$.*

*Proof.* Unfortunately, there are quite a few cases that have to be considered. Each case is defined by the intersection of the set $\{\{t-1, t-2\}, \{t-1, t-3\}, \{t-2, t-3\}\}$ with the two matchings; the cases are labeled by those of the edges that are contained in the two matchings. Due to symmetry via flipping the roles of $M_1$ and $M_2$ a couple of cases are omitted.

Let us clarify and recall a few things. The union of two perfect matchings on some vertex set is known to always give a disjoint union of cycles plus possibly a set of edges (the single edges are obtained when both matchings agree on the inclusion of some edges; we could regard them also as cycles of length two using a double edge). For the $\mathtt{shrink}$-operator it is only important which of the incident edges are in $M_1$ and $M_2$. For our subsequent arguments we also need to consider how the cycles in $M_1$ and $M_2$ traverse the elements $t-1$, $-2$, and $t-3$, and whether there are further connected components not containing any of these three elements.

We let $\alpha := \alpha_{M_1}$ and $\beta := \alpha_{M_2}$ be the corresponding functions for $M_1$ and $M_2$ that for any element return the one that it is matched to. Accordingly, the cycles in $M_1 \cup M_2$ will contain subpaths $(\alpha(t-1), t-1, \beta(t-1))$, $(\alpha(t-2), t-2, \beta(t-2))$, and $(\alpha(t-3), t-3, \beta(t-3))$. Depending on the edges incident on $t-1, t-2$, and $t-3$, we may get fewer paths: E.g., if $\{t-1, t-2\} \in M_1$ then $\alpha(t-1) = t-2$ and $\alpha(t-2) = t-1$; we get paths $(\beta(t-2), t-2, t-1, \beta(t-1))$ and $(\alpha(t-3), t-3, \beta(t-3))$.

The $\mathtt{shrink}$-operator will only change the edges with both endpoints in $\{t-1, t-2, t-3, \alpha(t-1), \alpha(t-2), \alpha(t-3), \beta(t-1), \beta(t-2), \beta(t-3)\}$, i.e., all other edges are preserved in the projection/image. Thus, the way that the above paths are connected into cycles in $M_1 \cup M_2$ is preserved in $M_1^0 \cup M_2^1$ and $M_1^1 \cup M_2^0$. In particular, we may assume that $M_1 \cup M_2$ contains no cycle disjoint from $t-1, t-2, t-3$ since it would be preserved too; in that case neither of the three pairs can form a Hamiltonian cycle and there is nothing left to show. Otherwise, clearly, the way that $\alpha(t-1), \alpha(t-2), \ldots, \beta(t-3)$ are connected affects directly whether $M_1 \cup M_2$ is a Hamiltonian cycle. For the other two pairs of matchings created by $\mathtt{shrink}$ these connections are the same but the $\mathtt{shrink}$-operator leads to different connections around $t-1, t-2$, and $t-3$ (in particular it removes $t-1$ and $t-2$), hence the total number of cycles may (and will) differ.

**1) $\{t-1, t-2\} \in M_1 \cap M_2$:** Clearly $M_1 \cup M_2$ is no Hamiltonian cycle as both matchings contain $\{t-1, t-2\}$. We observe that $M_1^1 = \mathtt{shrink}(M_1, 1) = \emptyset$ and $M_2^1 = \mathtt{shrink}(M_2, 1) = \emptyset$. Hence neither $M_1^0 \cup M_2^1$ nor $M_1^1 \cup M_2^0$ is a Hamiltonian cycle; this implies our claim.

**2) $\{t-1, t-3\} \in M_1 \cap M_2$:** Clearly $M_1 \cup M_2$ is no Hamiltonian cycle as both matchings contain $\{t-1, t-3\}$. We observe that $M_1^0 = \mathtt{shrink}(M_1, 0) = \emptyset$ and $M_2^0 = \mathtt{shrink}(M_2, 0) = \emptyset$. Hence neither $M_1^0 \cup M_2^1$ nor $M_1^1 \cup M_2^0$ is a Hamiltonian cycle; this implies our claim.

**3) $\{t-2, t-3\} \in M_1 \cap M_2$:** Clearly $M_1 \cup M_2$ is no Hamiltonian cycle as both matchings contain $\{t-2, t=3\}$. We will see that either both or none of $M_1^0 \cup M_2^1$ and $M_1^1 \cup M_2^0$ form a Hamiltonian cycle. By Definition A.1 we have

$$\mathtt{shrink}_t(M, b) := (M \setminus \{\{t-1, \alpha_M(t-1)\}, \{t-2, t-3\}\}) \cup \{\{t-3, \alpha_M(t-1)\}\},$$

when $\{t-2, t-3\} \in M$. Thus $M_1^0 = M_1^1$ and $M_2^1 = M_2^0$. It follows that $M_1^0 \cup M_2^1$ is a Hamiltonian cycle if and only if $M_1^1 \cup M_2^0$ is a Hamiltonian cycle, as needed.

**4) $\{t-1, t-2\} \in M_1$:** We observe that $M_1 \cup M_2$ contains subpaths $(\alpha(t-3), t-3, \beta(t-3))$ and $(\beta(t-1), t-1, t-2, \beta(t-2))$. By Definition A.1 we get

$$M_1^0 = M_1 \setminus \{\{t-1, t-2\}\},$$
$$M_2^1 = (M_2 \setminus \{\{t-1, \beta(t-1)\}, \{t-2, \beta(t-2)\}\}) \cup \{\{\beta(t-1), \beta(t-2)\}\},$$
$$M_1^1 = \emptyset,$$
$$M_2^0 = (M_2 \setminus \{\{t-1, \beta(t-1)\}, \{t-2, \beta(t-2)\}, \{t-3, \beta(t-3)\}\}) \cup$$
$$\cup \{\{t-3, \beta(t-2)\}, \{\beta(t-1), \beta(t-3)\}\}.$$

Hence, we know that $M_1^1 \cup M_2^0$ cannot form a Hamiltonian cycle (and in the following, we will omit $M_i^j$ if its partner is the empty matching). Further, we get that $M_1^0 \cup M_2^1$ contains subpaths $(\alpha(t-3), t-3, \beta(t-3))$ and $(\beta(t-1), \beta(t-2))$. Since these subpaths require the same configuration of the remaining vertices, as the subpaths of $M_1 \cup M_2$ we get that $M_1 \cup M_2$ is a Hamiltonian cycle if and only if $M_1^0 \cup M_2^1$ is a Hamiltonian cycle; this implies our claim.

**5) $\{t-1, t-3\} \in M_1$:** We observe that $M_1 \cup M_2$ contains subpaths $(\alpha(t-2), t-2, \beta(t-2))$ and $(\beta(t-1), t-1, t-3, \beta(t-3))$. By Definition A.1 we get

$$M_1^0 = \emptyset,$$
$$M_1^1 = (M_1 \setminus \{\{t-1, t-3\}, \{t-2, \alpha(t-2)\}\}) \cup \{\{t-3, \alpha(t-2)\}\},$$
$$M_2^0 = (M_2 \setminus \{\{t-1, \beta(t-1)\}, \{t-2, \beta(t-2)\}, \{t-3, \beta(t-3)\}\}) \cup$$
$$\cup \{\{t-3, \beta(t-2)\}, \{\beta(t-1), \beta(t-3)\}\}.$$

This time $M_1^0 \cup M_2^1$ cannot form a Hamiltonian cycle, since $M_1^0 = \emptyset$. For $M_1^1 \cup M_2^0$ we get subpaths $(\alpha(t-2), t-3, \beta(t-2))$ and $(\beta(t-1), \beta(t-3))$. Hence, the remaining edges complete $M_1 \cup M_2$ to a Hamiltonian cycle if and only if the same is true for $M_1^1 \cup M_2^0$.

**6) $\{t-2, t-3\} \in M_1$:** We observe that $M_1 \cup M_2$ contains subpaths $(\alpha(t-1), t-1, \beta(t-1))$ and $(\beta(t-2), t-2, t-3, \beta(t-3))$. By Definition A.1 we get

$$M_1^0 = (M_1 \setminus \{\{t-1, \alpha(t-1)\}, \{t-2, t-3\}\}) \cup \{\{t-3, \alpha(t-1)\}\},$$
$$M_2^1 = (M_2 \setminus \{\{t-1, \beta(t-1)\}, \{t-2, \beta(t-2)\}\}) \cup \{\{\beta(t-1), \beta(t-2)\}\},$$
$$M_1^1 = (M_1 \setminus \{\{t-1, \alpha(t-1)\}, \{t-2, t-3\}\}) \cup \{\{t-3, \alpha(t-1)\}\},$$
$$M_2^0 = (M_2 \setminus \{\{t-1, \beta(t-1)\}, \{t-2, \beta(t-2)\}, \{t-3, \beta(t-3)\}\}) \cup$$
$$\cup \{\{t-3, \beta(t-2)\}, \{\beta(t-1), \beta(t-3)\}\}.$$

Now, for $M_1^0 \cup M_2^1$ we get subpaths $(\alpha(t-1), t-3, \beta(t-3))$ and $(\beta(t-1), \beta(t-2))$, and for $M_1^1 \cup M_2^0$ we get subpaths $(\alpha(t-1), t-3, \beta(t-2))$ and $(\beta(t-1), \beta(t-3))$.

If the remaining edges contain any cycles, then none of $M_1 \cup M_2$, $M_1^0 \cup M_2^1$, and $M_1^1 \cup M_2^0$ form a Hamiltonian cycle (in the following we will tacitly ignore this case). Otherwise, all remaining edges form two vertex-disjoint paths that start and end in different vertices of $\{\alpha(t-1), \beta(t-1), \beta(t-2), \beta(t-3)\}$; for convenience, if, e.g., $\alpha(t-1) = \beta(t-2)$ then we take this as a trivial path from $\alpha(t-1)$ to $\beta(t-2)$. There are three ways in which those paths can connect the four vertices:

1. $(\alpha(t-1), \ldots, \beta(t-1))$ and $(\beta(t-2), \ldots, \beta(t-3))$: In this case $M_1 \cup M_2$ is no Hamiltonian cycle, but both $M_1^0 \cup M_2^1$ and $M_1^1 \cup M_2^0$ are Hamiltonian cycles. (This can be easily seen by combining the corresponding two subpaths with the two paths through the remaining edges: Either we get a single (Hamiltonian) cycle, or there are two cycles.)

2. $(\alpha(t-1), \ldots, \beta(t-2))$ and $(\beta(t-1), \ldots, \beta(t-3))$: In this case $M_1 \cup M_2$ and $M_1^0 \cup M_2^1$ are Hamiltonian cycles, and $M_1^1 \cup M_2^0$ is no Hamiltonian cycle.

3. $(\alpha(t-1), \ldots, \beta(t-3))$ and $(\beta(t-1), \ldots, \beta(t-2))$: In this case $M_1 \cup M_2$ and $M_1^1 \cup M_2^0$ are Hamiltonian cycles, and $M_1^0 \cup M_2^1$ is no Hamiltonian cycle.

Thus, for all configurations of the remaining edges we get that $M_1 \cup M_2$ forms a Hamiltonian cycle if and only if exactly one of $M_1^0 \cup M_2^1$ and $M_1^1 \cup M_2^0$ is a Hamiltonian cycle.

**7) $\{t-1, t-2\} \in M_1$ and $\{t-1, t-3\} \in M_2$:** We observe that $M_1 \cup M_2$ contains a subpath $(\alpha(t-3), t-3, t-1, t-2, \beta(t-2))$. By Definition A.1 we get

$$M_1^0 = M_1 \setminus \{\{t-1, t-2\}\},$$
$$M_2^1 = (M_2 \setminus \{\{t-1, t-3\}, \{t-2, \beta(t-2)\}\}) \cup \{\{t-3, \beta(t-2)\}\},$$
$$M_1^1 = \emptyset,$$
$$M_2^0 = \emptyset.$$

Thus, $M_1^1 \cup M_2^0$ is no Hamiltonian cycle. For $M_1^0 \cup M_2^1$ we get a single subpath $(\alpha(t-3), t-3, \beta(t-2))$. It follows easily that $M_1 \cup M_2$ is a Hamiltonian cycle if and only if $M_1^0 \cup M_2^1$ is a Hamiltonian cycle.

**8) $\{t-1, t-2\} \in M_1$ and $\{t-2, t-3\} \in M_2$:** We observe that $M_1 \cup M_2$ contains a subpath $(\alpha(t-3), t-3, t-2, t-1, \beta(t-1))$. By Definition A.1 we get

$$M_1^0 = M_1 \setminus \{\{t-1, t-2\}\},$$
$$M_2^1 = (M_2 \setminus \{\{t-1, \beta(t-1)\}, \{t-2, t-3\}\}) \cup \{\{t-3, \beta(t-1)\}\},$$
$$M_1^1 = \emptyset.$$

Thus $M_1^1 \cup M_2^0$ is no Hamiltonian cycle. For $M_1^0 \cup M_2^1$ we get a single subpath $(\alpha(t-3), t-3, \beta(t-1))$. This implies that $M_1 \cup M_2$ is a Hamiltonian cycle if and only if $M_1^0 \cup M_2^1$ is a Hamiltonian cycle.

**9) $\{t-1, t-3\} \in M_1$ and $\{t-2, t-3\} \in M_2$:** We observe that $M_1 \cup M_2$ contains a subpath (
$alpha(t-2), t-2, t-3, t-1, \beta(t-1))$. By Definition A.1 we get

$$M_1^0 = \emptyset,$$
$$M_1^1 = (M_1 \setminus \{\{t-1, t-3\}, \{t-2, \alpha(t-2)\}\}) \cup \{\{t-3, \alpha(t-2)\}\},$$
$$M_2^0 = (M_2 \setminus \{\{t-1, \beta(t-1)\}, \{t-2, t-3\}\}) \cup \{\{t-3, \beta(t-1)\}\}.$$

Thus $M_1^0 \cup M_2^1$ is no Hamiltonian cycle. For $M_1^1 \cup M_2^0$ we get a single subpath $(\alpha(t-2), t-3, \beta(t-1))$. This implies that $M_1 \cup M_2$ is a Hamiltonian cycle if and only if $M_1^1 \cup M_2^0$ is a Hamiltonian cycle.

**10) $\{\{t-1, t-2\}, \{t-1, t-3\}, \{t-2, t-3\}\} \cap (M_1 \cup M_2) = \emptyset$:** In this final case we can only observe the trivial subpaths in $M_1 \cup M_2$, namely $(\alpha(t-1), t-1, \beta(t-1))$, $(\alpha(t-$

$2), t - 2, \beta(t - 2))$, and $(\alpha(t - 3), t - 3, \beta(t - 3))$. By Definition A.1 we get

$$\begin{aligned}
M_1^0 &= (M_1 \setminus \{\{t-1, \alpha(t-1)\}, \{t-2, \alpha(t-2)\}, \{t-3, \alpha(t-3)\}\}) \cup \\
&\quad \cup \{\{t-3, \alpha(t-2)\}, \{\alpha(t-1), \alpha(t-3)\}\}, \\
M_2^1 &= (M_2 \setminus \{\{t-1, \beta(t-1)\}, \{t-2, \beta(t-2)\}\}) \cup \{\{\beta(t-1), \beta(t-2)\}\}, \\
M_1^1 &= (M_1 \setminus \{\{t-1, \alpha(t-1)\}, \{t-2, \alpha(t-2)\}\}) \cup \{\{\alpha(t-1), \alpha(t-2)\}\}, \\
M_2^0 &= (M_2 \setminus \{\{t-1, \beta(t-1)\}, \{t-2, \beta(t-2)\}, \{t-3, \beta(t-3)\}\}) \cup \\
&\quad \cup \{\{t-3, \beta(t-2)\}, \{\beta(t-1), \beta(t-3)\}\}.
\end{aligned}$$

Thus, $M_1^0 \cup M_2^1$ has subpaths

$$(\alpha(t-1), \alpha(t-3)), (\alpha(t-2), t-3, \beta(t-3)), (\beta(t-1), \beta(t-2)),$$

and $M_1^1 \cup M_2^0$ has subpaths

$$(\alpha(t-1), \alpha(t-2)), (\alpha(t-3), t-3, \beta(t-2)), (\beta(t-1), \beta(t-3)).$$

Thus, whether any of the three unions of perfect matchings is a Hamiltonian cycle depends again on the paths given by all remaining edges. As before, if those edges give rise to cycles disjoint from $t-1, t-2, t-3$ then none of the three unions is a Hamiltonian cycle and we are done. Else, the remaining edges serve exactly to create three vertex disjoint paths between the vertices $\alpha(t-1), \alpha(t-2), \alpha(t-3), \beta(t-1), \beta(t-2)$, and $\beta(t-3)$; there are 15 configurations of such paths (i.e., partitions of this set of endpoints into three pairs of connected vertices). Note that again we treat cases like $\alpha(t-1) = \beta(t-2)$ has a path connecting $\alpha(t-1)$ and $\beta(t-2)$.

It can be easily verified that for each of the 15 configurations we get the desired property that $M_1 \cup M_2$ is a Hamiltonian cycle if and only if exactly one of $M_1^0 \cup M_2^1$ and $M_1^1 \cup M_2^0$ is a Hamiltonian cycle. For completeness, we provide a table containing all 15 cases; see Table 1. This completes the proof. □

As a special case of Lemma A.3 when the second matching is in our intended basis, we get the following lemma. Both lemmas are used for proving the main theorem of this section.

**Lemma A.4.** *Let $t$ be an even integer $t \geq 6$ and let $M \in \Pi_2(U_t)$. Furthermore let $X = X(t, ab)$ where $b \in \{0, 1\}$ and $a$ denotes a $0, 1$-string of length $\frac{t}{2} - 2$. Then $M \cup X$ is a Hamiltonian cycle if and only if $\mathtt{shrink}(M, b) \cup X(t-2, a)$ is a Hamiltonian cycle, i.e.,*

$$[M \sqcap X(t, ab) = \{U_t\}] = [\mathtt{shrink}(M, b) \sqcap X(t-2, a) = \{U_{t-2}\}].$$

We are now set up to prove Theorem 3.4; let us recall the theorem statement first.

**Theorem A.5.** *3.4 Let $t \geq 2$ be an even integer and let $M_1, M_2 \in \Pi_2(U_t)$. It holds that*

$$[M_1 \sqcap M_2 = \{U_t\}] \equiv \sum_{a \in \{0,1\}^{t/2-1}} [M_1 \sqcap X(t, a) = \{U_t\}] \cdot [M_2 \sqcap X(t, \overline{a}) = \{U_t\}],$$

*where $X(t, a), X(t, \overline{a}) \in \mathbf{X}_t$ according to Definition 3.1. (Each matching in $\mathbf{X}_t$ occurs exactly twice, once as $X(t, a)$ and once as $X(t, \overline{a})$.)*

We establish a factorization of the Matching Connectivity matrix as the product of two smaller rectangular matrices: The first matrix has rows labeled by (all) perfect matchings and columns labeled by basis matchings $X(t, a)$. The second matrix has rows labeled basis matchings $X(t, \overline{a})$ and columns labeled by (all) perfect matchings. The entries of both rectangular matrices are as in the large one: we have 1 if the two matchings form a Hamiltonian cycle and 0 else (so in fact they are submatrices of the large matrix). Note that the two perfect matchings from $\mathbf{X}_t$ in each summand are exactly the unique pairs that give Hamiltonian cycles (among basis matchings). Also note that the main statement of the theorem is given in arithmetic modulo two.

| configuration of remaining edges as paths between $\alpha(t-1), \alpha(t-2), \alpha(t-3), \beta(t-1), \beta(t-2), \beta(t-3)$ | $M_1 \cup M_2$ is H-cycle | $M_1^0 \cup M_2^1$ is H-cycle | $M_1^1 \cup M_2^0$ is H-cycle |
|---|---|---|---|
| $\alpha(t-1)$-$\alpha(t-2)$, $\alpha(t-3)$-$\beta(t-1)$, $\beta(t-2)$-$\beta(t-3)$ | 1 | 1 | 0 |
| $\alpha(t-1)$-$\alpha(t-2)$, $\alpha(t-3)$-$\beta(t-2)$, $\beta(t-1)$-$\beta(t-3)$ | 1 | 1 | 0 |
| $\alpha(t-1)$-$\alpha(t-2)$, $\alpha(t-3)$-$\beta(t-3)$, $\beta(t-1)$-$\beta(t-2)$ | 0 | 0 | 0 |
| $\alpha(t-1)$-$\alpha(t-3)$, $\alpha(t-2)$-$\beta(t-1)$, $\beta(t-2)$-$\beta(t-3)$ | 1 | 0 | 1 |
| $\alpha(t-1)$-$\alpha(t-3)$, $\alpha(t-2)$-$\beta(t-2)$, $\beta(t-1)$-$\beta(t-3)$ | 0 | 0 | 0 |
| $\alpha(t-1)$-$\alpha(t-3)$, $\alpha(t-2)$-$\beta(t-3)$, $\beta(t-1)$-$\beta(t-2)$ | 1 | 0 | 1 |
| $\alpha(t-1)$-$\beta(t-1)$, $\alpha(t-2)$-$\alpha(t-3)$, $\beta(t-2)$-$\beta(t-3)$ | 0 | 1 | 1 |
| $\alpha(t-1)$-$\beta(t-1)$, $\alpha(t-2)$-$\beta(t-2)$, $\alpha(t-3)$-$\beta(t-3)$ | 0 | 1 | 1 |
| $\alpha(t-1)$-$\beta(t-1)$, $\alpha(t-2)$-$\beta(t-3)$, $\alpha(t-3)$-$\beta(t-2)$ | 0 | 0 | 0 |
| $\alpha(t-1)$-$\beta(t-2)$, $\alpha(t-2)$-$\alpha(t-3)$, $\beta(t-1)$-$\beta(t-3)$ | 1 | 1 | 0 |
| $\alpha(t-1)$-$\beta(t-2)$, $\alpha(t-2)$-$\beta(t-1)$, $\alpha(t-3)$-$\beta(t-3)$ | 0 | 1 | 1 |
| $\alpha(t-1)$-$\beta(t-2)$, $\alpha(t-2)$-$\beta(t-3)$, $\alpha(t-3)$-$\beta(t-1)$ | 1 | 0 | 1 |
| $\alpha(t-1)$-$\beta(t-3)$, $\alpha(t-2)$-$\alpha(t-3)$, $\beta(t-1)$-$\beta(t-2)$ | 1 | 0 | 1 |
| $\alpha(t-1)$-$\beta(t-3)$, $\alpha(t-2)$-$\beta(t-1)$, $\alpha(t-3)$-$\beta(t-2)$ | 1 | 1 | 0 |
| $\alpha(t-1)$-$\beta(t-3)$, $\alpha(t-2)$-$\beta(t-2)$, $\alpha(t-3)$-$\beta(t-1)$ | 0 | 1 | 1 |

Table 1: All 15 configurations for paths connecting the vertices $\alpha(t-1), \alpha(t-2), \alpha(t-3), \beta(t-1), \beta(t-2)$, and $\beta(t-3)$, and their effect on whether $M_1 \cup M_2$, $M_1^0 \cup M_2^1$, and $M_1^1 \cup M_2^0$ form a Hamiltonian cycle (denoted by 1 and 0 otherwise.

*Proof.* Let us quickly verify the theorem for the base case $t = 2$. There is a unique perfect matching $\{\{0,1\}\}$ for $U_2 = \{0,1\}$. Thus, we can only pick $M_1 = M_2 = \{\{0,1\}\}$. Hence, for the left hand side of the theorem statement we get

$$LHS = [M_1 \sqcap M_2 = \{U_2\}] = 1.$$

Regarding the right hand side the only choice for $a \in \{0,1\}^{t/2-1}$ in the sum is $a = \varepsilon$ (of course $\overline{\varepsilon} = \varepsilon$). We recall that $X(2,\varepsilon) = \{\{0,1\}\}$, thus we get

$$RHS = [M_1 \sqcap X(2,\varepsilon) = \{U_2\}] \cdot [M_2 \sqcap X(2,\varepsilon) = \{U_2\}] = 1.$$

This completes the base case $t = 2$.

Now, we will perform an inductive argument to show that the theorem holds also for all even $t \geq 4$. We start from the right hand side of the theorem statement and apply Lemmas A.3 and A.4 to transform it into the left hand side. As a first step, we split the right hand side sum into two parts, depending on the last position of $a$ (to do so formally we shorten $a$ by one position and explicitly specify 0 or 1 to be appended).

$$RHS = \sum_{a \in \{0,1\}^{t/2-2}} [M_1 \sqcap X(t, a0) = \{U_t\}] \cdot [M_2 \sqcap X(t, \overline{a0}) = \{U_t\}]$$

$$+ \sum_{a \in \{0,1\}^{t/2-2}} [M_1 \sqcap X(t, a1) = \{U_t\}] \cdot [M_2 \sqcap X(t, \overline{a1}) = \{U_t\}]$$

Note that $\overline{a0} = \overline{a}1$ and $\overline{a1} = \overline{a}0$. Let us recall the statement of Lemma A.4:

$$[M \sqcap X(t, ab) = \{U_t\}] = [\texttt{shrink}_t(M, b) \sqcap X(t-2, a) = \{U_{t-2}\}].$$

Using the shorthand $M_i^j := \texttt{shrink}_t(M_i, j)$ we can thus rewrite the RHS as follows.

$$RHS = \sum_{a \in \{0,1\}^{t/2-2}} [M_1^0 \sqcap X(t-2, a) = \{U_{t-2}\}] \cdot [M_2^1 \sqcap X(t-2, \overline{a}) = \{U_{t-2}\}]$$
$$+ \sum_{a \in \{0,1\}^{t/2-2}} [M_1^1 \sqcap X(t-2, a) = \{U_{t-2}\}] \cdot [M_2^0 \sqcap X(t-2, \overline{a}) = \{U_{t-2}\}]$$

We note that both sums go over all bitstrings $a$ of length $\frac{t}{2} - 2 = \frac{t-2}{2} - 1$. Thus, after brief inspection of the summands, we see that we can apply the inductive assumption to both sums since $M_i^j \in \Pi_2(U_{t-2})$. We obtain

$$RHS = [M_1^0 \sqcap M_2^1 = \{U_{t-2}\}] + [M_1^1 \sqcap M_2^0 = \{U_{t-2}\}].$$

Finally, by a direct application of Lemma A.3 we obtain

$$RHS = [M_1 \sqcap M_2 = \{U_t\}] = LHS,$$

which is what we intended to prove. Thus, the theorem holds for all even integer $t \geq 2$, as claimed. □